%% file: CosmoSimMargConverg.astro-ph.2.tex
\documentclass[usegraphicx,usenatbib,useAMS]{mn2e} 
\usepackage{amsfonts}
\usepackage{amssymb}
\usepackage{amsmath}
\usepackage{bm}

\input{defs.tex}

\title[Cosmological constraints and $N$-body codes]
{Precision cosmology in muddy waters: Cosmological constraints and $N$-body codes}

\author[{\it R.~E.~Smith et al. }] 
       {\parbox{\textwidth}{
           Robert~E.~Smith$^{1,2,3}$\thanks{res@mpa-garching.mpg.de}, 
           Darren~S.~Reed$^{2}$, 
           Doug Potter$^{2}$, 
           Laura Marian$^{3}$,\\ 
           Martin Crocce$^{4}$ \& 
           Ben Moore$^{2}$}
         \vspace*{8pt} \\ 
         $^1$ Max-Planck Institut f\"ur Astrophysik, 
           Karl Schwarzschild Str.1, D-85741,
           Garching bei M\"unchen, Germany\\
         $^2$ Institute for Theoretical Physics, University of Zurich, Zurich CH 8037\\ 
         $^3$ Argelander-Institute for Astronomy, Auf dem H\"ugel 71, D-53121 Bonn, Germany\\
         $^4$ Institut de Ci\`encies de l'Espai (IEEC-CSIC),  E-08193 Bellaterra (Barcelona), Spain }
       
\begin{document}

\maketitle



\begin{abstract}
Future large-scale structure surveys of the Universe will aim to
constrain the cosmological model and the true nature of dark energy
with unprecedented accuracy. In order for these surveys to achieve
their designed goals, they will require predictions for the nonlinear
matter power spectrum to sub-percent accuracy. Through the use of a
large ensemble of cosmological $N$-body simulations, we demonstrate
that if we do not understand the uncertainties associated with
simulating structure formation, i.e. knowledge of the `optimal'
simulation parameters, and simply seek to marginalize over them, then
the constraining power of such future surveys can be significantly
reduced. However, for the parameters $\{n_s,h,\Omega_b,\Omega_m\}$,
this effect can be largely mitigated by adding the information from a
CMB experiment, like Planck. In contrast, for the amplitude of
fluctuations $\sigma_8$ and the time-evolving equation of state of
dark energy $\{w_0,w_a\}$, the mitigation is mild. On marginalizing
over the simulation parameters, we find that the dark-energy figure of
merit can be degraded by $\sim2$. This is likely an optimistic
assessment, since we do not take into account other important
simulation parameters. A caveat is our assumption that the Hessian of
the likelihood function does not vary significantly when moving from
our adopted to the {\em optimal} simulation parameter set. This paper
therefore provides strong motivation for rigorous convergence testing
of $N$-body codes to meet the future challenges of precision
cosmology.
\end{abstract}




\section{Introduction}

Future spectro/imaging surveys of the low-redshift Universe such as
DES\footnote{www.darkenergysurvey.org/},
KiDS\footnote{www.astro-wise.org/projects/KIDS/}
Euclid\footnote{sci.esa.int/euclid} and
WFIRST\footnote{wfirst.gsfc.nasa.gov/} will aim to constrain the
cosmological model to unprecedented accuracy. This will require
impressive handling of every step of the observational pipeline in
order to limit the possibility of systematic errors that may degrade
the constraints on cosmological parameters. Besides the observational
processing, there will also be a similarly high demand placed on our
ability to generate theoretical predictions that are sufficiently
accurate not to bias inferred cosmological parameters. These
predictions will take the form of a set of estimators for the primary
observables that we intend to measure and their covariance, and also
most likely their cross-covariance. The latter will be required for
robustly testing for modifications to gravity \citep{Reyesetal2010}.

In galaxy clustering or cosmic shear surveys of the Universe, the
primary observables of interest are related to the matter power
spectrum and its evolution with redshift. The matter power spectrum is
the two-point covariance of the matter fluctuations transformed into
Fourier space. The power spectrum provides a wealth of information on
the cosmological parameters \citep{Dodelson2003,Weinberg2008}.  In
order to maximize the amount of information obtainable from the power
spectrum, we need to understand its dependence on the cosmological
parameters in the nonlinear regime.

A number of theoretical and semi-empirical techniques are available
for predicting the nonlinear power spectrum: such as the halo model
\citep{Seljak2000,PeacockSmith2000,MaFry2000a,Smithetal2003};
perturbation theory \citep{Bernardeauetal2002,CrocceScoccimarro2008};
and scale transformations
\citep{Hamiltonetal1991,PeacockDodds1996}. However, it is widely
acknowledged that $N$-body simulations provide the most direct path
towards this answer. Cosmological $N$-body simulations are not without
pit-falls \citep{Heitmannetal2005, Heitmannetal2008, Reedetal2012}, and
in general they depend on a number of pseudo-free parameters, such as:
the number of particles used to represent the phase-space; the
box-size; the redshift at which the initial conditions are given; the
scale on which two-body forces are softened, etc.  If a particle-mesh
(PM) approach is employed then one additionally needs to set the scale
above which forces are solved using mesh based techniques. If a tree
technique is used, then one additionally needs to adopt a choice for
the multipole order to which forces are expanded and the type and
depth of the tree to be used. If both are used, then one also needs to
set parameters that interpolate between the tree and PM methods.

Given the complexity of the state-of-the-art $N$-body codes, we are
then lead to ask the following questions:
\begin{itemize}

\item How do we determine the values of the {\em optimal} simulation parameters?

\item How much would forecasted parameter constraints be degraded if
  we were to marginalize over the simulation parameters?

\item How does this affect the dark energy `figure of merit'?

\end{itemize}

In this paper we shall employ a large ensemble of $N$-body simulations
to directly answer these latter two questions, and leave the first for 
future study.

The paper can be broken down as follows: In \S\ref{sec:likelihood} we
provide a brief overview of the necessary theoretical background, and
in particular we give a brief review of the Fisher matrix approach to
forecasting cosmological constraints.  In \S\ref{sec:simulations} we
describe the large ensemble of simulations that we employ. In
\S\ref{sec:fiducial} we present results concerning the fiducial model
power spectrum and its covariance matrix. In \S\ref{sec:variations} we
explore the dependence of the matter power spectrum on cosmological
and simulation parameters. In \S\ref{sec:Fisher} we use the Fisher
matrix approach to explore how various assumptions concerning our
understanding of $N$-body simulations can impact the cosmological
parameter forecasts from future large-scale-structure surveys. In
\S\ref{sec:DE} we focus on the question of constraining the time
evolution of dark energy and how ignorance of simulation parameters
impacts the figure of merit. Finally, in \S\ref{sec:discussion} and
\S\ref{sec:conclusions} we discuss this approach, summarize our
findings and conclude.


\section{Forecasting cosmological constraints}\label{sec:likelihood}

\subsection{The {\em Gemeinsam} likelihood function}

\def\Vu{V_{\rm \mu}}

We are interested in forecasting the ability of a future survey of the
universe to constrain the cosmological parameter space. We may assess
this using the Fisher matrix approach.

Consider a particular statistic that we will estimate from the survey
data, and let us be concrete and take this to be the matter power
spectrum $P(k)$. The power spectrum may be defined as
\citep{Peebles1980}:
\be 
\Vu\left<\delta(\bk_1)\delta(\bk_2)\right>\equiv 
P(k_1)\delta^{K}_{\bk_1,-\bk_2}\ , \label{eq:Pk}
\ee
where the Fourier modes of the density field are given by
\be
\tilde{\delta}(\bk) =  \frac{1}{\Vu}\int \dx \exp\left[-i\bk\cdot\bx\right] \delta(\bx) 
\ee
and where the over-density field is defined as:
\be \delta(\bx)\equiv\frac{\rho(\bx)-\rhob}{\rhob} \ ,\ee
with $\rho$ and $\rhob$ being the local and background density.

A given theoretical cosmological power spectrum depends on the
wavenumber $k$ -- here we focus on the real-space isotropic function
-- and also the cosmological parameters $\bm\theta$. We are also
interested in the case where the theoretical predictions also depend
on a set of internal simulation parameters $\bm\psi$. Let us write the
augmented vector of cosmological and simulation parameters as
$\bm\phi=(\bm\theta,\bm\psi)$. We denote the measurement of
$P(k)$ at wavenumber $k_i$ by $P_i$ and the theoretical
(simulated) spectra by $P^{\rm sim}(k_i|\bm\phi)$. Notice that here we
are making the approximation that $P_i$ does not depend on $\theta$;
this in fact is not true and any measurement of $P$ requires the
assumption of a cosmological model (we reserve further discussion of
this for future work and note that this simply makes the analysis
sub-optimal).

Let us adopt a Bayesian approach to the analysis of our data and write
the $m$ measurements of the power spectra at wavenumbers \mbox{${\bf
    k}\rightarrow \{k_1,\dots,k_m\}$}, as \mbox{${\rm \bf
    P}\rightarrow \{P_1,\dots,P_m\}$}. The probability that our survey
yields observations ${\rm \bf P}$, given the cosmological and
simulation parameters $\bm\phi$, is $L({\rm \bf P}|{\bm\phi})$ -- the
{\em likelihood}. If the likelihood is Gaussian, then we have
\be 
L({\bf P}|\bm\phi)=\frac{1}{(2\pi)^{n/2}|{\rm \bf C}|^{1/2}}
\exp\left[-\frac{1}{2} y_i(\bm\phi) C^{-1}_{ij}(\bm\phi) y_j(\bm\phi)\right]
\label{eq:like1}\ee
where we have made use of the Einstein summation convention. In the
above equation we also defined
\be y_i(\bm\phi)\equiv P_i-\overline{P}^{\rm
  theory}(k_i|\bm\phi) \ , \ee
where $\overline{P}^{\rm theory}(k_i|\bm\phi)$ is the expectation
for the theory power spectrum.  ${\rm \bf C}(\bm\phi)$ is the
covariance matrix, which may be defined as:
\be C_{ij}    \equiv 
\left<\left[P_i-\overline{P}^{\rm theory}(k_i) \right]\left[P_j-\overline{P}^{\rm theory}(k_j) \right]\right> 
%
\ee
and $|\CC|$ is the determinant of the covariance matrix.  

Using Bayes theorem, the likelihood is directly related to the {\em
  posterior} probability, $p(\bm\phi|{\rm \bf P})$, through a set of 
{\em priors}, $p(\bm\phi)$, and is normalized by the {\em
  evidence}, $p({\rm \bf P})$:
\be p(\bm\phi|{\rm \bf P}) =
\frac{p(\bm\phi)L({\rm \bf P}|\bm\phi)}{p({\rm \bf P})} 
=  \frac{p(\bm\phi)L({\rm \bf P}|\bm\phi)}
{\int d {\bm\phi} p({\bm\phi}) L({\rm \bf P}|{\bm\phi})} \ .\
\ee
If the priors are flat, then the posterior probability is simply
proportional to the likelihood.  Close to its maximum, at $\bm\phi_0$,
we may Taylor expand the logarithm of the posterior, and for flat
priors also the log likelihood (${\mathcal L}\equiv \ln L$), to
obtain:
\ba \ln p(\bm\phi|{\rm \bf P}) & \propto & {\mathcal L}({\rm \bf P}|\bm\phi) \nn \\
& \approx & {\mathcal L}({\rm \bf P}|\bm\phi_0)-
\frac{1}{2}\Delta\phi_{\alpha}{\mathcal H}_{\alpha\beta}(\bm\phi_0)
\Delta\phi_{\beta} +\dots \ ,\ea
where in the above:
$\Delta\bm\phi\equiv\left(\bm\phi-\bm\phi_{0}\right)$ are deviations
of the parameters from the fiducial values; the first derivative
vanished at the maximum; the second derivative is identified as
\be {\mathcal H}_{\alpha\beta}\equiv- \nabla^{\phi}_\alpha\nabla^{\phi}_\beta {\mathcal L} \ee
and it is given the name of Hessian, or curvature matrix. We have also
used the notation:
$\nabla^{\phi}_\alpha\equiv\partial/\partial\phi_\alpha$. In
truncating this expression for the posterior at second order we are
implicitly assuming that the likelihood is also Gaussian in the
parameters. Hence, we may rewrite the above expression for the
posterior as,
\ba p(\bm\phi|{\rm \bf P}) & \approx & 
\frac{p(\bm\phi)}{p({\rm \bf P})} L(\bm\phi_0)\exp\!\left[-
\frac{1}{2}  \Delta\phi_\alpha {\mathcal H}_{\alpha\beta}(\bm\phi_0)
\Delta\phi_\beta \right]
\ .\ea
Thus ${\mathcal H}_{\alpha\beta}$ informs us about errors on the parameters and
how different parameters may be correlated with respect to each other
-- in the context of their effects on the data.


Since the likelihood itself depends on the data, it is also a random
variable. Taking an ensemble average over many realizations of the
data, we arrive at the Fisher matrix:
\be {\mathcal F}_{\alpha\beta}=\left<{\mathcal
    H}_{\alpha\beta}\right>=-\left<\frac{\partial^2 \ln
    L}{\partial\phi_\alpha\partial\phi_\beta}\right> \ .\ee

Considering the division into cosmological and simulation parameters,
this matrix may be written schematically as:
\be {\mathcal F}^{\phi\phi}=
\left(
\begin{array}{ll}
{\mathcal F}^{\theta\theta} & {\mathcal F}^{\theta\psi} \\
{\mathcal F}^{\psi\theta} & {\mathcal F}^{\psi\psi}
\end{array}
\right) \ .
\ee
where ${\mathcal F}^{\theta\theta}$, ${\mathcal F}^{\theta\psi}$ and
${\mathcal F}^{\psi\psi}$ denote the Fisher matrices of the
cosmological, cosmological-cross-simulation and simulation parameter
spaces, respectively.

From the Fisher matrix one can obtain the expected marginalized error
on parameter $\phi_i$ and the covariance between parameters
$(\phi_i,\phi_j)$:
\be 
\sigma_{ii}\ge\sqrt{\left[ {\mathcal F}^{\phi\phi}\right]^{-1}_{ii}}\ ; \hspace{1cm}
\sigma_{ij}\ge\sqrt{\left[ {\mathcal F}^{\phi\phi}\right]^{-1}_{ij}}\ . \label{eq:FishCos}
\ee
We can also obtain conditional errors for the cosmological parameters,
conditioned on the simulation parameters possessing a particular
value:
\be 
\sigma_{ii}\ge\sqrt{\left[ {\mathcal F}^{\theta\theta}\right]^{-1}_{ii}}\ ; \hspace{1cm}
\sigma_{ij}\ge\sqrt{\left[ {\mathcal F}^{\theta\theta}\right]^{-1}_{ij}}\ . \label{eq:FishCosSim}
\ee
These expressions represent the minimum variance bounds (MVB) 
\citep[for a derivation see][]{Heavens2009}.

Lastly, we note that for the specific case of a Gaussian likelihood,
it can be shown that the Fisher matrix takes on the special form
\citep{Tegmarketal1997,Heavens2009}:
\ba {\mathcal F}_{\alpha\beta} =  \frac{1}{2} {\rm Tr}
\left[\CC^{-1}\CC_{,\alpha}\CC^{-1}\CC_{,\beta}\right] +
\left[{\rm \bf P}^{\rm theory}_{,\alpha}\right]^{T}{\rm\bf C}^{-1}{\rm \bf P}^{\rm theory}_{,\beta}
\label{eq:Fisher}.
\ea

Our first objective may now be reformulated as the following
questions:

\begin{itemize}
\item How do the MVBs obtained from ${\mathcal F}^{\theta\theta}$
  compare with those for ${\mathcal F}^{\phi\phi}$? How do our
  parameter forecasts degrade when we marginalize over simulation
  parameters?
\end{itemize}

\begin{table}
  \caption{Cosmological parameters used for the fiducial {\tt
      zHORIZON} suite of simulations and the variations with respect
    to the cosmological parameters. The columns are: name of
    simulation series; density parameters for matter, dark energy and
    baryons; the equation of state parameter for the dark energy
    $P_{w}=w\rho_{\rm w}$; normalization and primordial spectral index
    of the power spectrum; dimensionless Hubble parameter $h$,
    respectively.
    \label{tab:cospar}}
\centering{
\begin{tabular}{c|ccccccccccc}
\hline 
Param. & $\Omega_m$ &  $\Omega_b$ & $w_0$  &  $w_a$ &
$\sigma_8$  & $n$ &  $h$ \\
\hline
{\tt Fid.} & 0.25       & 0.04        &  -1.0        & 0.0 &  0.8        & 1.00 & 0.7  \\
\hline
{\tt V1}  & {\bf 0.20} & 0.04        &  -1.0        & 0.0 &  0.8        & 1.00 & 0.7  \\
{\tt V2}  & {\bf 0.30} & 0.04        &  -1.0        & 0.0 &  0.8        & 1.00 & 0.7  \\
\hline
{\tt V3}  & 0.25       & {\bf 0.035} &  -1.0        & 0.0 &  0.8        & 1.00 & 0.7 \\
{\tt V4} & 0.25       & {\bf 0.045} &  -1.0        & 0.0 &  0.8        & 1.00 & 0.7 \\
\hline
{\tt V5}  & 0.25       & 0.04        &  {\bf -1.2}  & 0.0 &  0.8        & 1.00 & 0.7 \\
{\tt V6}  & 0.25       & 0.04        &  {\bf -0.8}  & 0.0 &  0.8        & 1.00 & 0.7 \\
\hline
{\tt V7} & 0.25       & 0.04        &  -1.0        & {\bf -0.1} &  0.8        & 1.00 & 0.7 \\
{\tt V8} & 0.25       & 0.04        &  -1.0        & {\bf 0.1} &  0.8        & 1.00 & 0.7 \\
\hline
{\tt V9}  & 0.25       & 0.04        &  -1.0        & 0.0 &  {\bf 0.7}  & 1.00 & 0.7  \\
{\tt V10}  & 0.25       & 0.04        &  -1.0        & 0.0 &  {\bf 0.9}  & 1.00 & 0.7  \\
\hline
{\tt V11}  & 0.25       & 0.04        &  -1.0        & 0.0 & 0.8        & {\bf 0.95} & 0.7 \\
{\tt V12}  & 0.25       & 0.04        &  -1.0        & 0.0 & 0.8        & {\bf 1.05} & 0.7 \\
\hline
{\tt V13} & 0.25       & 0.04        &  -1.0        & 0.0 &  0.8        & 1.00 & {\bf 0.65} \\
{\tt V14} & 0.25       & 0.04        &  -1.0        & 0.0 &  0.8        & 1.00 & {\bf 0.75} 
\end{tabular}}
\vspace{0.3cm}
  \caption{The {\tt Gadget-2} parameters used for all fiducial
    simulations, and the variations with respect to the simulation
    parameters. The columns are: name of simulation series; simulation
    parameter varied; fiducial value; simulated variations.
\label{tab:simpar}}
\begin{tabular}{lllll} 
\hline
Simulation & Parameters & Fiducial & Low & High  \\
\hline
{\tt S1/S2} & {\tt ErrTolForceAcc}  & 0.005 & 0.004 & 0.006 \\     
{\tt S3/S4} & {\tt ErrTolIntAcc}    & 0.025 & 0.02 & 0.03 \\
{\tt S5/S6} & {\tt ErrTolTheta}     & 0.45  & 0.4 & 0.5 \\
{\tt S7/S8} & {\tt PMGRID}          & 750   & 500 & 1000 \\
{\tt S9/S10} & {\tt MaxRMSDispFac}   & 0.2   & 0.15 & 0.25 \\
{\tt S11/S12} & {\tt Softening}       & 0.03  & 0.025 & 0.035\\
{\tt S13/S14} & {\tt RCUT}            & 4.5   & 4.0 & 5.0 \\ 
{\tt S15/S16} & {\tt ASMTH}           & 1.25  & 1.15 & 1.35 \\
{\tt S17/S18} & {\tt MaxSizeTstep}    & 0.025 & 0.020 & 0.03 \\
\hline
\end{tabular}
\end{table}


\section{The $N$-body Simulations}\label{sec:simulations}


In order to study the Fisher information we have generated a large
suite of $N$-body simulations. As \Eqn{eq:Fisher} demonstrates, one
needs to compute the derivatives of the theoretical power spectra with
respect to the parameters $\bm\phi$ and also the covariance matrix. In
fact, one also needs the derivatives of the covariance matrix with
respect to the parameters $\bm\phi$. Since estimating the covariance
matrix is a sufficiently challenging task in itself, we shall reserve
the inclusion of information from this for future study. Henceforth,
we shall drop the first term in \Eqn{eq:Fisher} from our analysis
\citep[for further justification of this approximation
  see][]{Tegmark1997}.

In order to determine the covariance matrix we have simulated 200
realizations of our fiducial cosmological model. The specific
cosmological parameters that we adopted are for a flat $\Lambda$CDM
model with:
$\{\sigma_8=0.8,\Omega_m=0.25,\Omega_b=0.04,w_0=-1.0,w_a=0.0,h=0.7,n_s=1.0\}
$ where: $\sigma_8$ is the variance of mass fluctuations in a top-hat
sphere of radius $R=8 \Mpc$; $\Omega_m$ and $\Omega_b$ are the matter
and baryon density parameters; $w_0$ and $w_a$ are the constant and
time-evolving equation-of-state parameters for the dark energy,
i.e. $P_{\rm DE}/\rho_{\rm DE}\equiv w(a)=w_0+(1-a)w_a$; $h$ is the
dimensionless Hubble parameter; and $n$ is the power-law index of the
primordial density power spectrum. Our adopted values were inspired by
the results of the WMAP experiment
\citep{Komatsuetal2009}. Table~\ref{tab:cospar} contains further
details of the cosmological parameters of the fiducial model.

All of the $N$-body simulations were run on the {\tt zBOX-3} and {\tt
  Schr\"odinger} supercomputers at the University of Z\"urich, using
the publicly available Tree-PM code {\tt GADGET-2}
\citep{Springel2005}, with a slight modification that permitted a
time-evolving equation of state for dark energy, specified by the
parameters $\{w_0,w_a\}$.  This code was used to follow with
high-force accuracy the nonlinear evolution under gravity of
\mbox{$N=750^3$} equal mass particles in a comoving cube of length
\mbox{$L=1500\Mpc$}, giving a total sample volume of order
\mbox{$V\sim540 \Gpccube$}. Newtonian two-body forces are softened
below scales \mbox{$l_{\rm soft}=60\, \kpc$}. We shall refer to this
suite of simulations as the {\tt zHORIZON} runs (Z\"urich Horizon
simulations).  The transfer function for the simulations was generated
using the publicly available {\tt cmbfast} code
\citep{SeljakZaldarriaga1996}, with high sampling of the spatial
frequencies on large scales. For the time evolving dark energy models
we used the code {\tt CAMB} \citep{Lewisetal1999} and with the dark
energy module of \citet{HuSawicki2007}.
 
Initial conditions were lain down at redshift $z=49$ using the serial
version of the publicly available {\tt 2LPT} code
\citep{Crocceetal2006}.  The {\tt zBOX-3} runs took roughly
$\sim$20Hrs per run on 256 cores, and the {\tt Schr\"odinger} runs
took $\sim$6Hrs per run on 256 cores.  For all of the realizations
snapshots were output at a number of redshifts, though for this study
we focus only on the results at $z=\{1, 0.5, 0\}$.  For
completeness, the {\tt Gadget-2} parameters that we used are presented
in Table~\ref{tab:simpar}.

In order to evaluate the derivatives of the power spectrum with
respect to the cosmological parameters, we have performed an
additional 56 simulations -- the cosmological variations, labeled
V1-V14. We have considered the effect of changing a single
cosmological parameter, whilst holding the remaining parameters
fixed. For each such modification we ran 4 simulations. We used
double-sided variations, e.g.  $P(k|\bm\phi+{\Delta}\phi_i)$ and
$P(k|\bm\phi-{\Delta}\phi_i)$, as this enables more accurate
computations of the numerical derivatives, which will be important for
our Fisher-matrix estimates. Also, in order to reduce the noise in
these estimates, we matched the initial Gaussian random field of each
realization with the corresponding one from the fiducial model. Full
details of these simulations are summarized in Table \ref{tab:cospar}.

To estimate the derivatives of the power spectrum with respect to the
simulation parameters, we have performed another 18 simulations -- the
simulation variations, labeled S1-S18. This time we keep the
cosmological parameters as the fiducial ones and explore the effect of
changing a single simulation parameter, whilst holding the remaining
ones fixed. For each such modification we ran a single simulation, but
again we considered double-sided variations, with matched initial
Gaussian random fields so as to decrease the noise when estimating
derivatives. The exact list of simulation parameters that we have
sampled are presented in Table~\ref{tab:simpar}.


\begin{figure}
\centering{
  \includegraphics[width=8cm,clip=]{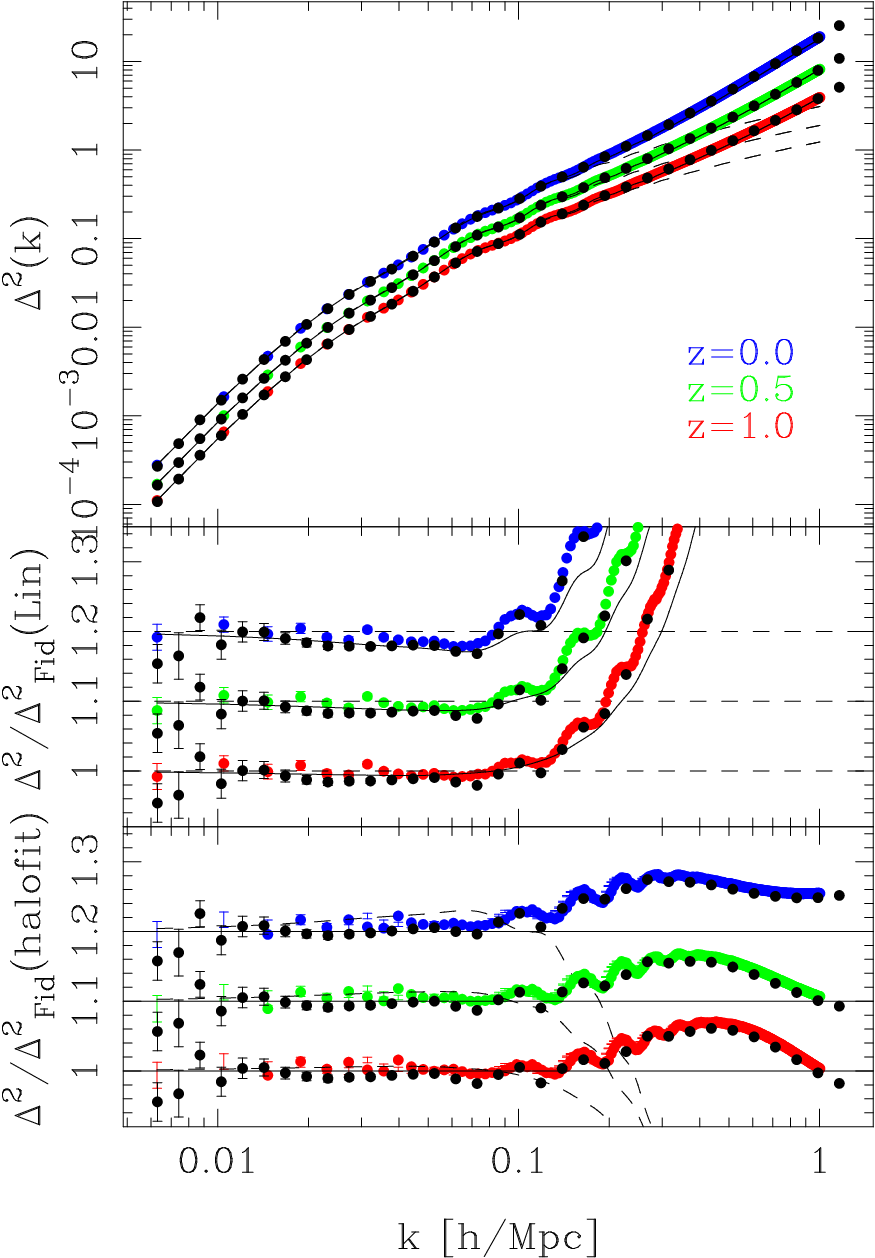}}
\caption{\small{The ensemble-averaged dark matter power spectrum for the
    200 fiducial realizations, at redshifts $z\in\{1,0.5,0\}$, with errors
    per realization. {\em Top panel:} absolute dimensionless power,
    $\Delta^2=k^3P(k)/2\pi^2$. {\em Central panel:} the ratio of the
    mean power spectra with respect to the linear theory. {\em Bottom
      panel:} ratio of power spectra with respect to the nonlinear
    predictions from the fitting formula {\tt halofit}
    \citep{Smithetal2003}. In all panels the solid and dashed lines
    denote the nonlinear and linear theory predictions,
    respectively. Note that in the central and bottom panels, the
    $z=0.5$ and $z=0$ results have been off-set by 0.1 and 0.2 in
    the vertical direction for clarity. }\label{fig:FidPow}}
\end{figure}


\section{Analysis I: The fiducial model}\label{sec:fiducial}


\subsection{Power spectrum}

As a first exploration of the simulation data we compute the matter
power spectrum at the redshifts of interest.

The power spectrum in the simulation cube for a given Fourier mode is
as described in \Eqn{eq:Pk}. In practice, the power is estimated by
averaging over all wavemodes in thin spherical shells in $k$-space --
band-powers. The band-power-averaged power spectrum can be written,
\ba \widehat{P}^{d}(k_i) & = & 
\frac{\Vu}{V_{s,i}}\int_{V_{s,i}}\dk\,
\left<\delta^{d}(\bk)\delta^{d}(-\bk)\right> \nn \\
& = &
\frac{\Vu}{N_{k_i}}\sum_{j=1}^{N_{k_i}}
\left<\delta^{d}(\bk_j)\delta^{d}(-\bk_j)\right>\ ,
\label{eq:powest}\ea
where the average is over the k-space shell $V_s$, of volume
\be V_{s,i}=\int_{k_i-\Delta k/2}^{k_i+\Delta k/2}\dk
=4\pi k_i^2\Delta k\left[1+\frac{1}{12}
\left(\frac{\Delta k}{k_i}\right)^2\right]\ \ee
and where $N_{k_i}=V_{s,i}/V_k$ is the total number of modes in the
shell.  $V_k=k_f^3$ is the fundamental $k$-space cell volume and
$k_f=2\pi/L$ is the fundamental wavemode. 

Notice that in \Eqn{eq:powest} we have used the superscript $d$, this
stands for discrete, since we make a Fourier decomposition of the
point-sampled field. For a point-sampled process, the power spectrum
is related to that of the continuous mass density field through the
relation \citep{Peebles1980,Smith2009}:
\be P^d(k)=P^c(k)+\frac{1}{\nbar} \ , \label{eq:autospectrum}\ee
where $P^c$ is the power spectrum of the underlying continuous
field. The constant term on the right-hand side of the equation is
more commonly referred to as the `shot-noise correction' term, and is
the additional variance introduced through discreteness, where
$\nbar=N/\Vu$. However, we do not make such a correction, since the
initial particle configuration of the simulation is not strictly a
point sampling of a continuous density field. In particular, for the
grid starts that we use, applying the above correction at early times
and on large scales would lead to negative power. However, on small
scales the discreteness correction is well described by
\Eqn{eq:autospectrum}. We therefore make no discreteness correction,
but use the form of \Eqn{eq:autospectrum} to judge when such effects
are significant \citep[for more discussion
  see][]{Smithetal2003}.

In order to compute the power spectrum, we apply the standard Fast Fourier
Transform (FFT)-based approach \citep[see for
  example][]{Smithetal2003,Jing2005,Smithetal2008b}. We use a
`cloud-in-cell' (CIC) mass assignment scheme for the simulation
particles, and deconvolve each Fourier mode accordingly with the window
function. The power spectrum estimator is then given by
\Eqn{eq:powest}. We use FFT grids with $N_{\rm grid}=1024$ cells per
dimension, and this sets the minimum and maximum spatial frequencies
to: $k_{\rm min}=2\pi/L=0.0042\kMpc$ and $k_{\rm Ny}=\pi N_{\rm
  grid}/L=2.15 \kMpc$. In practice, the power on length scales
$k>k_{\rm NY}$ will get `aliased' to larger spatial scales, and so we
take $k_{\rm max}=k_{\rm Ny}/2$. In this study we have decided to
estimate the power spectrum in 35 logarithmically spaced band powers
in the interval \mbox{$k\in[0.0042,1.0]\kMpc$}. We adopt this strategy
in order to obtain sufficiently high signal-to-noise estimates of the
covariance matrix.


\begin{figure}
\centering{
  \includegraphics[width=8cm,clip=]{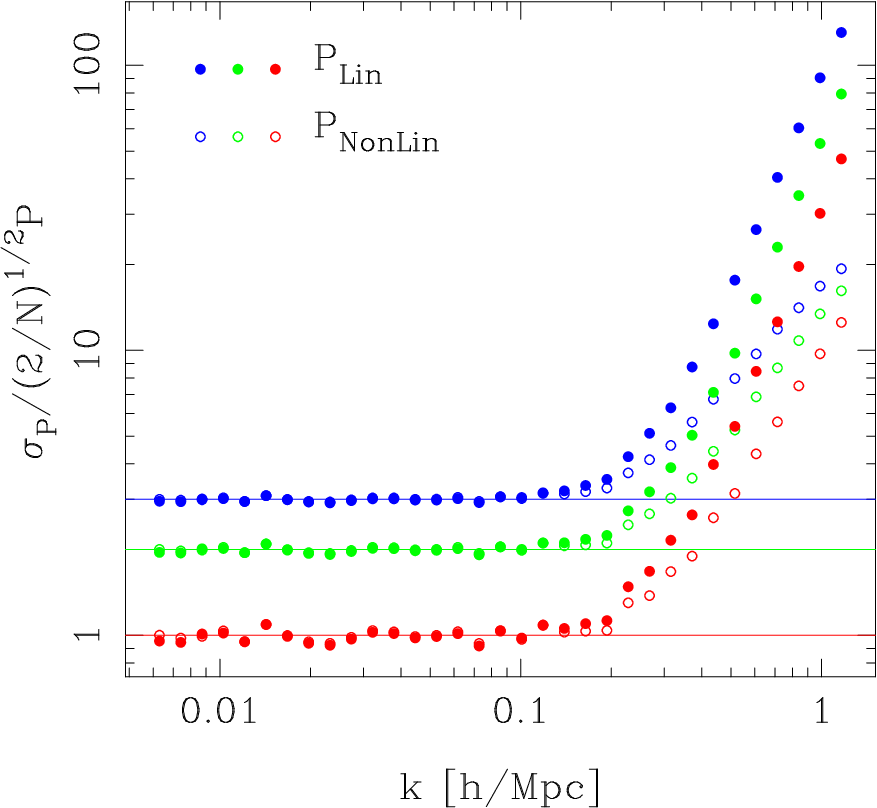}}
\caption{\small{Ratio of the measured error in the power spectrum to
    the Gaussian-predicted error, i.e. Eq.~\ref{eq:CovGauss}. The
    measured error is obtained from the ensemble of 200 $N$-body
    simulations. Blue, green and red points depict the results for
    redshifts $z\in\{0, 0.5, 1\}$, respectively. The solid and open
    symbols show the results obtained when the linear theory and
    measured nonlinear power spectra are used to make the predictions
    from the Gaussian error formula. For clarity, the $z=0$ and
    $z=0.5$ results have been off-set by factors of 3 and 2 on the
    $y$-axis, respectively.}\label{fig:FidPowVar}}
\end{figure}


Figure \ref{fig:FidPow} shows the ensemble-averaged dark matter power
spectrum for the 200 realizations, at the redshifts
$z\in\{1,0.5,0\}$, denoted by the green, red and blue points,
respectively. The coloured points actually show the power spectra
obtained from a linear binning, where the bin spacing is in units of
the fundamental $k$-cell spacing, $k_{f}$. The black points denote the
results for the 35 logarithmically spaced bins, and the error bars are
on the mean. In the top panel of the figure we show the dimensionless
power, which may be defined:
\be \Delta^2(k)\equiv\frac{4\pi}{(2\pi)^3}k^3P(k)\ .\ee
In the middle panel we show the ratio of the measured power spectra
with respect to the input linear theory power spectra. For clarity, we
offset the power spectra at $z=0$ by 20\% and at $z=0.5$ by 10\% in
the vertical direction.  The black solid line shows the nonlinear
power spectra predictions from {\tt halofit} \citep{Smithetal2003}.
The plot demonstrates that strong nonlinear amplification occurs on
scales $k\gtrsim 0.1\kMpc$, and that linear theory is not a good
approximation on these scales. In addition, nonlinear amplification is
not significantly weaker at higher redshifts.  The bottom panel shows
the ratio of the measured power spectra with respect to {\tt
  halofit}. Again we have offset the power spectra for clarity. {\tt
  halofit} is able to describe the measured power spectra to better
than 10\% on the scales investigated. The BAO wiggles appear
emphasized when one takes the ratio of the nonlinear spectrum with the
linear one. As was explained in \citep{GuzikBernsteinSmith2007}, this
is due to the fact that the BAO in the nonlinear spectrum are damped
and smoothed relative to linear, and so when one takes the ratio with
the linear one sees stronger acoustic oscillations.

The shot-noise correction to the power spectrum is $P_{\rm
  shot}=1/\nbar=8\Mpccube$, which in terms of the dimensionless power
is $\Delta^2_{\rm shot}=4k^3/\pi^2$. At $k\sim1\kMpc$, this is
$\Delta^2_{\rm shot}\sim 0.4$. Thus for $z=1$, the power has roughly a
10\% correction, which by $z=0$ is reduced to 2\%.


\begin{figure*}
\centering{
  \includegraphics[width=10.5cm,angle=-90.0,clip=]{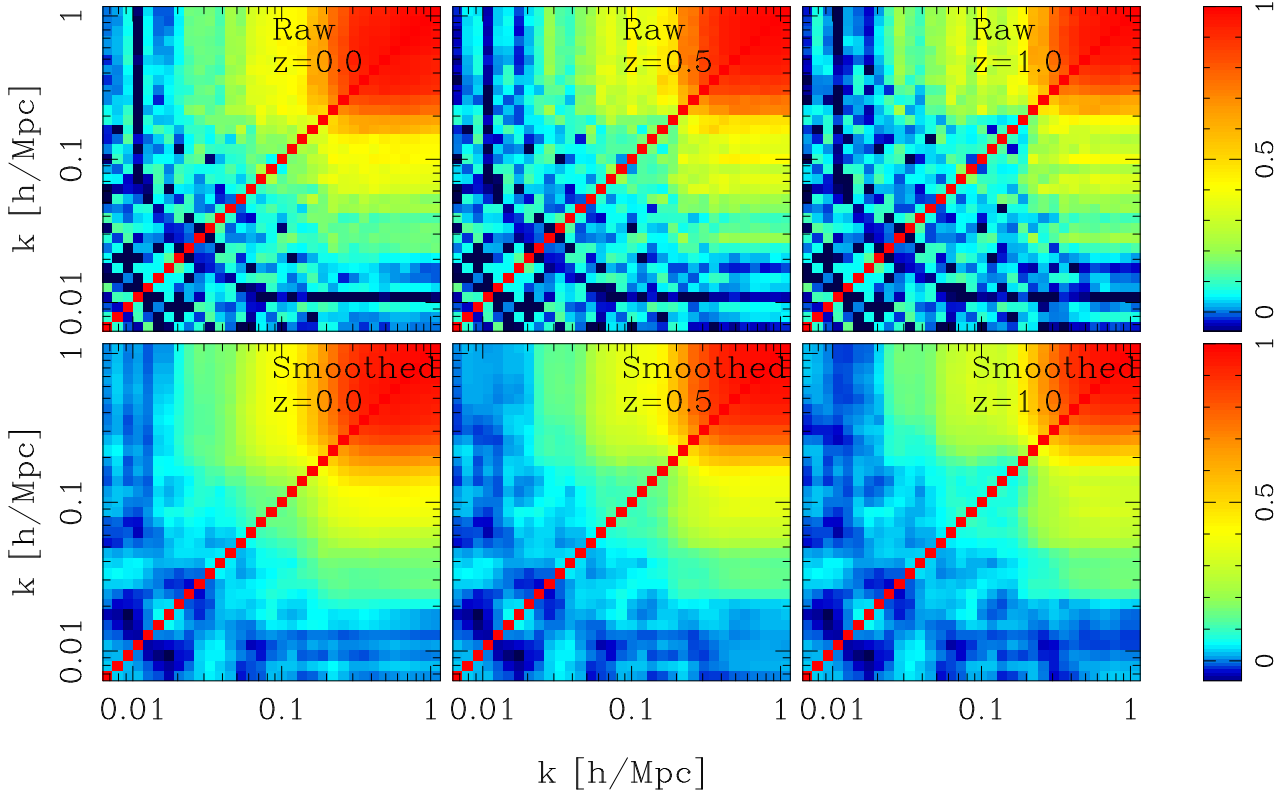}}
\caption{\small{{\em Top panel}: Evolution of the power spectrum
    correlation matrix as a function of wavenumber, estimated from the
    ensemble of 200 simulations. The left, central and right panels
    show the results for $z=\{1.0,\,0.5,\,0.0\}$, respectively. Bottom
    panel: same as above, only the correlation matrix has been box-car
    smoothed by a square top-hat filter of size $3\times3$
    pixels. This clearly reduces the noise in the correlation matrix
    on large scales. }\label{fig:FidPowCov}}
\end{figure*}


\subsection{Covariance matrix}\label{ssec:cov}

An unbiased estimator for the covariance between different band power
estimates can be obtained through:
\ba
\widehat{C}_{ij} & = & \frac{1}{N_{\rm E}-1}
\sum_{\alpha=1}^{N_{\rm E}}
\left[\widehat{P}_i^{(\alpha)}-\widehat{\overline{P}}_i\right]
\left[\widehat{P}_j^{(\alpha)}-\widehat{\overline{P}}_j\right]\ ;  \\
\widehat{\overline{P}}_i & = & \frac{1}{N_{\rm E}}\sum_{\alpha=1}^{N_{\rm
      E}}\widehat{P}_i^{(\alpha)} ,
\ea
where $N_{\rm E}$ is the number of realizations. 

Following \citet{Scoccimarroetal1999b} and \citet{Smith2009}, a
theoretical expression for the bin-averaged covariance matrix of the
matter power spectrum, obtained from a set of densely-sampled tracers
of the mass field, can be written:
\be
{C}_{ij} =  \frac{{T}_{ij}}{\Vu}  
+\frac{2}{N_{k}}\left[\widehat{\overline{P}}_i\right]^2 
\delta^{K}_{i,j}  \label{eq:CovFull}\ ,
\ee
where ${T}_{ij}$ is the shell-averaged, connected part of the
trispectrum in parallelogram configuration:
\be {T}_{ij}
\equiv 
\int \frac{\dk_1}{V_{s,i}}\frac{\dk_2}{V_{s,j}} \tilde{T}(\bk_1,\bk_2,-\bk_1,-\bk_2) \ ,
\ee
with
$\tilde{T}(\bk_1,\bk_2,\bk_3,\bk_4)\Vu^3\equiv\left<\delta(\bk_1)\dots\delta(\bk_4)\right>\delta^{\rm
  K}_{\bk_1+\dots+\bk_4,{\bf 0}}$ being the matter trispectrum. Note
that for a Gaussian random field the connected part of the trispectrum
vanishes, i.e. $\tilde{T}=0$, and the covariance reduces to:
\be
{C}_{ij} = \frac{2}{N_{k_i}}\left[\widehat{\overline{P}}_i\right]^2 
\delta^{K}_{i,j}  \label{eq:CovGauss}\ .
\ee

Figure~\ref{fig:FidPowVar} shows the standard deviation for the matter
power spectrum, i.e. $C_{ii}^{1/2}$, measured from the simulations,
scaled in units of the square root of the Gaussian expectation for the
variance given in the equation above. The figure reveals that for
$k<0.1\kMpc$ the diagonal errors are reasonably well described by
\Eqn{eq:CovGauss}. However, on smaller scales we find that the errors
are significantly larger than one would expect from simple mode
counting. If one uses \Eqn{eq:CovGauss} with $P$ measured from the
simulations (open points in \Fig{fig:FidPowVar}), then the errors
appear to increase $\propto k$. This would suggest that in the deeply
nonlinear regime $T_{ii}\propto P^2_i$ as $k\rightarrow \infty$. This
scaling is consistent with the predictions from the 1-Loop
perturbation theory \citep{Scoccimarroetal1999b}.


\begin{figure*}
\centering{
  \includegraphics[width=9.2cm,angle=-90,clip=]{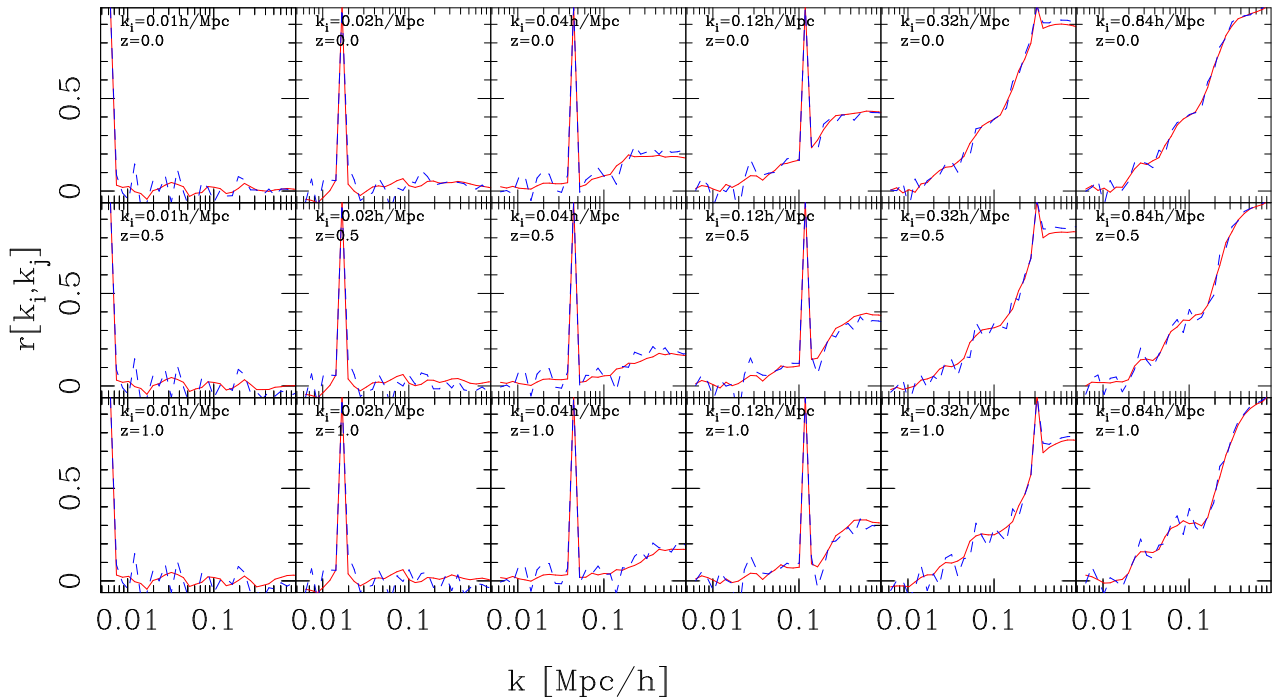}}
\caption{\small{Different rows of the power spectrum correlation
    matrix as a function of wavenumber. From left to right the columns
    show results for $k_1=\{0.01,0.02,0.04,0.12,0.32,0.84\}\kMpc$, as a
    function of $k_2$. The top, middle and bottom rows show the
    results for $z=\{0,0.5,1\}$, respectively. The dashed and
    solid lines show the estimates for the raw and box-car smoothed
    correlation matrix. }\label{fig:FidPowCovSlice}}
\end{figure*}


An interesting way to present the information in the off-diagonal
elements of the covariance matrix is through the cross-correlation
matrix. This may be defined as:
\be
r_{ij}\equiv\frac{C_{ij}}{\sqrt{C_{ii}C_{jj}}}\ 
\label{eq:corrmatrix} ,\ee
and it gives the strength of the covariance in a particular element
relative to the square root of the product of variances in the
relevant bins. It is therefore bound to the interval $r\in[-1,1]$.

The top panels in Figure~\ref{fig:FidPowCov} present the evolution of
the correlation matrix as a function of redshift from $z=1$ to
$z=0$. The increasingly redder/bluer colours demonstrate
increasing/decreasing correlation strength. The results on large
scales appear to be slightly noisy. The bottom panels of
Fig.~\ref{fig:FidPowCov} presents the same information, only here we
have performed a box-car smoothing of the correlation matrix in order
to reduce the noise. For each pixel, we take the average of all pixels
that are within 1-pixel from the current pixel centre, excluding the
pixels on the diagonal and being careful in our treatment of the edges
of the matrix. We also keep the diagonal elements fixed at $r=1$
\citep[for further discussion of this approach
  see][]{Mandelbaumetal2012}. This noise reduction strategy
constitutes a plausible alternative to various other \emph{ad-hoc}
approaches presented elsewhere in the literature \citep{Nganetal2012}.

For the case of both the raw and the noise-reduced matrix, the
off-diagonal correlations are in general non-zero and positive. The
correlations increase as the wavenumbers of the two considered band
powers are increased. Also, the correlation increases with decreasing
redshift. For our choice of binning and simulation volume, we find
that different power spectral band powers are $>50\%$ correlated for
$\{k_i,k_j\}\gtrsim 0.2\kMpc$ at $z=0$, and for $\{k_i,k_j\}\gtrsim
0.25\kMpc$ by $z=1$.

Figure~\ref{fig:FidPowCovSlice} presents slices through the power
spectrum correlation matrices measured at $z\in\{0,0.5,1\}$ for the
200 realizations of the fiducial model. These results point to a
reasonably good agreement between the raw and box-car-smoothed
covariance matrices.

The covariance matrix of the matter power spectrum has recently been
explored by \citet{Takahashietal2009} who ran 5000 PM simulations in
boxes of size $L=1\Gpc$ with $N=256^3$ particles. These simulations
are not of sufficiently-high spatial resolution to probe the
covariance of power spectrum estimates beyond scales of the order
$k\sim0.2(0.4)\kMpc$ at $\sim1(3)\%$ precision. Moreover, whilst they
did employ the more accurate 2LPT initial conditions -- as does our
study -- they also used a rather low start redshift of $z=20$, which
may induce small scale inaccuracies \citep{Reedetal2012}. On comparing
their results with ours, we note that whilst they have a factor of 25
times more simulations, each of our simulations has a factor of 3
times more volume. This makes the overall difference roughly a factor
of $\sim3$ in terms of $(S/N)$. We should therefore be able to obtain
a reasonably accurate covariance matrix. This is further mitigated by
our smoothing of the correlation matrix.

We also compare our study with that of \citet{Nganetal2012}, who used
the code {\tt CUBEP3M} to run 1000 simulations of boxes with
$L=600\Mpc$ and with $N=256^3$ particles. We underline that for this
choice of simulation set-up, the shot-noise corrections to the power
spectrum at $k=1\kMpc$ are 6\% and 30\% at $z=0$ and $z=1$,
respectively. Again, whilst their study used 1000 simulations, our
simulations have roughly 15 times more volume per run. Moreover, they
have explored the covariance matrix for 54 bins, nearly a factor of 2
times more than we employ, hence the relative statistical power of our
study should be at the very least comparable with their work. It is
also worth pointing out that \citet{Nganetal2012} found a 20\%
anti-correlation of band-powers on the largest scales. We find no
evidence of such a strong anti-correlation. We also note that the
power spectra from the simulations of \citet{Nganetal2012} appear to
show a worrying $\sim5-7\%$ positive off-set from the linear theory
predictions for the power spectrum on scales comparable to the
box-scale, although this issue may now be resolved
\citep{Harnois-Derapsetal2012}.

A more recent study by \citet{dePutteretal2012} has used a suite of
1024 simulations of $L=600\Mpc$ boxes and 160 simulations of a
$L=2400\Mpc$ box to explore the covariance matrix. They found that the
results from the large-box simulations were in reasonably good
agreement with the larger ensemble of smaller box simulations. 

Thus our results are in broad agreement with all of these works and
non-trivial band power correlations must be accounted for in
cosmological analysis of large-scale structure data.


\begin{figure*}
\centering{
\includegraphics[width=10cm,height=13cm,angle=-90.0,clip=]{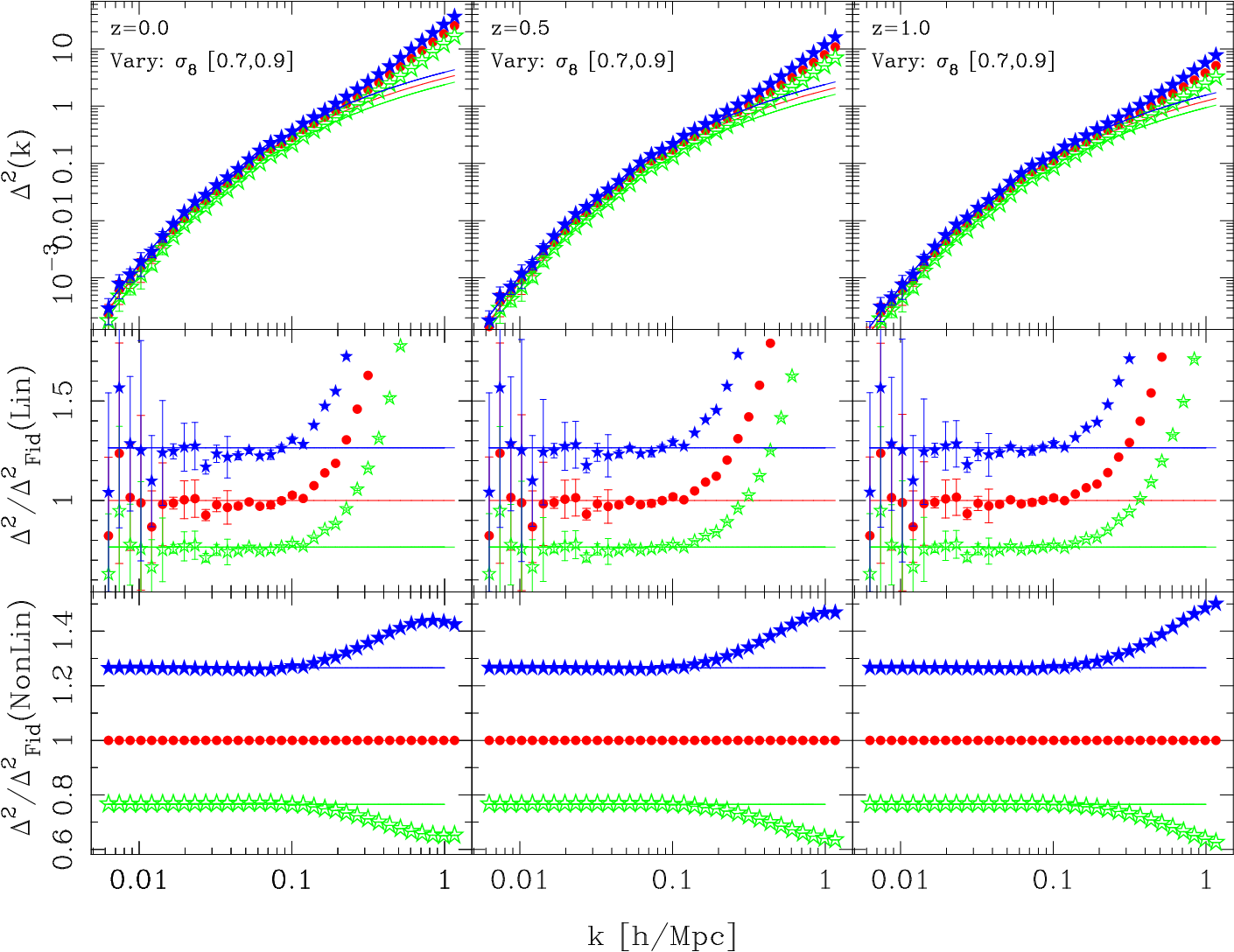}}
\caption{\small{Dependence of the nonlinear matter power spectrum on
    the power spectrum normalization parameter $\sigma_8$. In all
    panels the solid blue and open green stars depict the estimates from
    the $N$-body simulations with $\sigma_8=\{0.9,0.7\}$, and the
    solid red points denote the results for the fiducial model
    $\sigma_8=0.8$. The top panels show the absolute power spectrum;
    the central panels show the ratio of the spectra with respect to
    the fiducial linear theory predictions; the bottom panels show the
    ratio of the spectra with respect to the fiducial power
    spectrum. From left to right, the three columns represent results
    for epochs $z=\{0,0.5,1\}$, respectively.
\label{fig:PowVar1}}}
\vspace{0.5cm}
\centering{
\includegraphics[width=10cm,height=13cm,angle=-90.0,clip=]{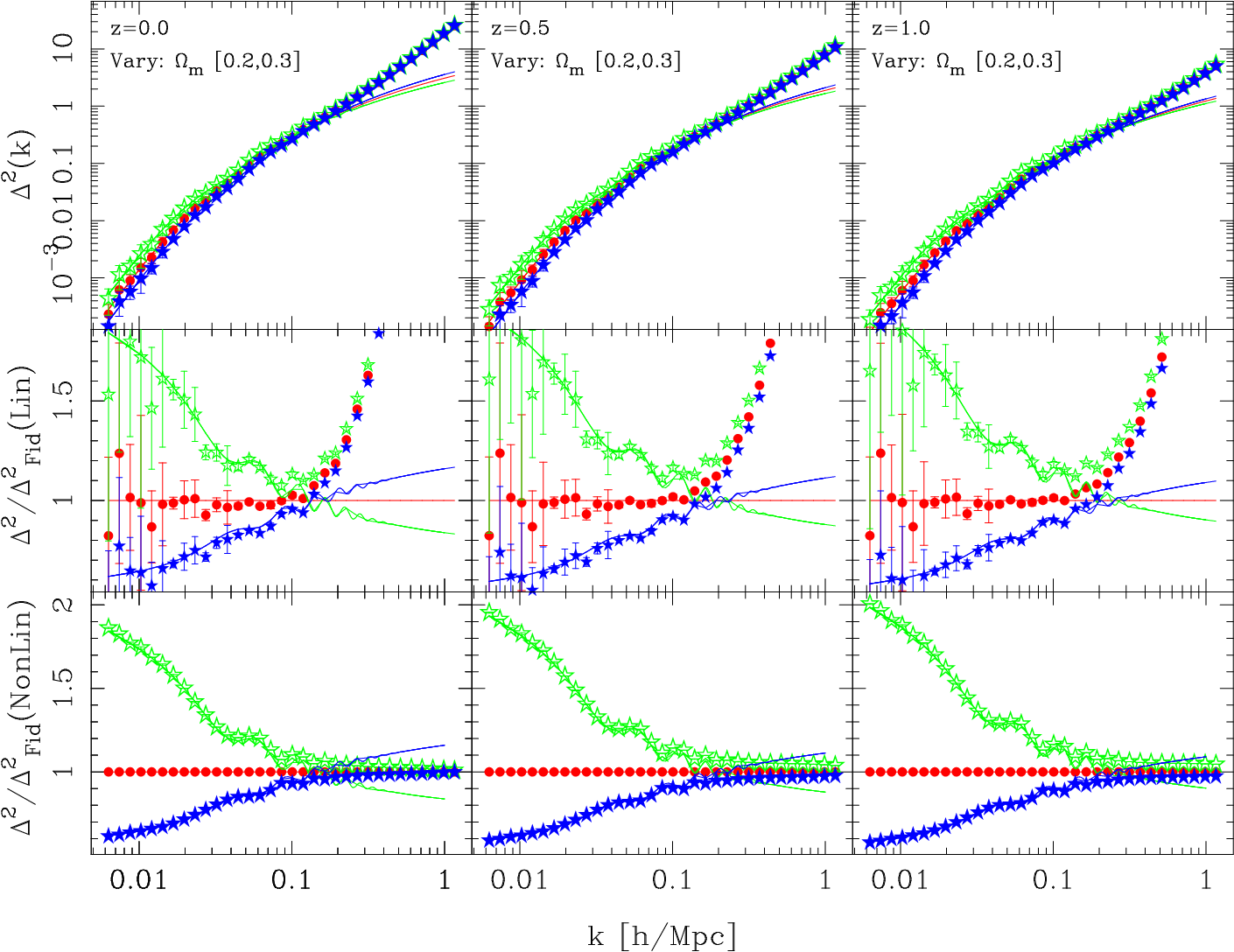}} 
\caption{\small{The same as \Fig{fig:PowVar1}, only this time showing
    the dependence of the power spectrum on the matter density parameter
    $\Omega_m$. The solid blue and open green stars depict the results
    for $\Omega_m=\{0.3,0.2\}$, and the solid red points denote the
    results for the fiducial model $\Omega_m=0.25$.
    \label{fig:PowVar2}}}
\end{figure*} 


\begin{figure*}
\centering{
\includegraphics[width=10cm,height=13cm,angle=-90.0,clip=]{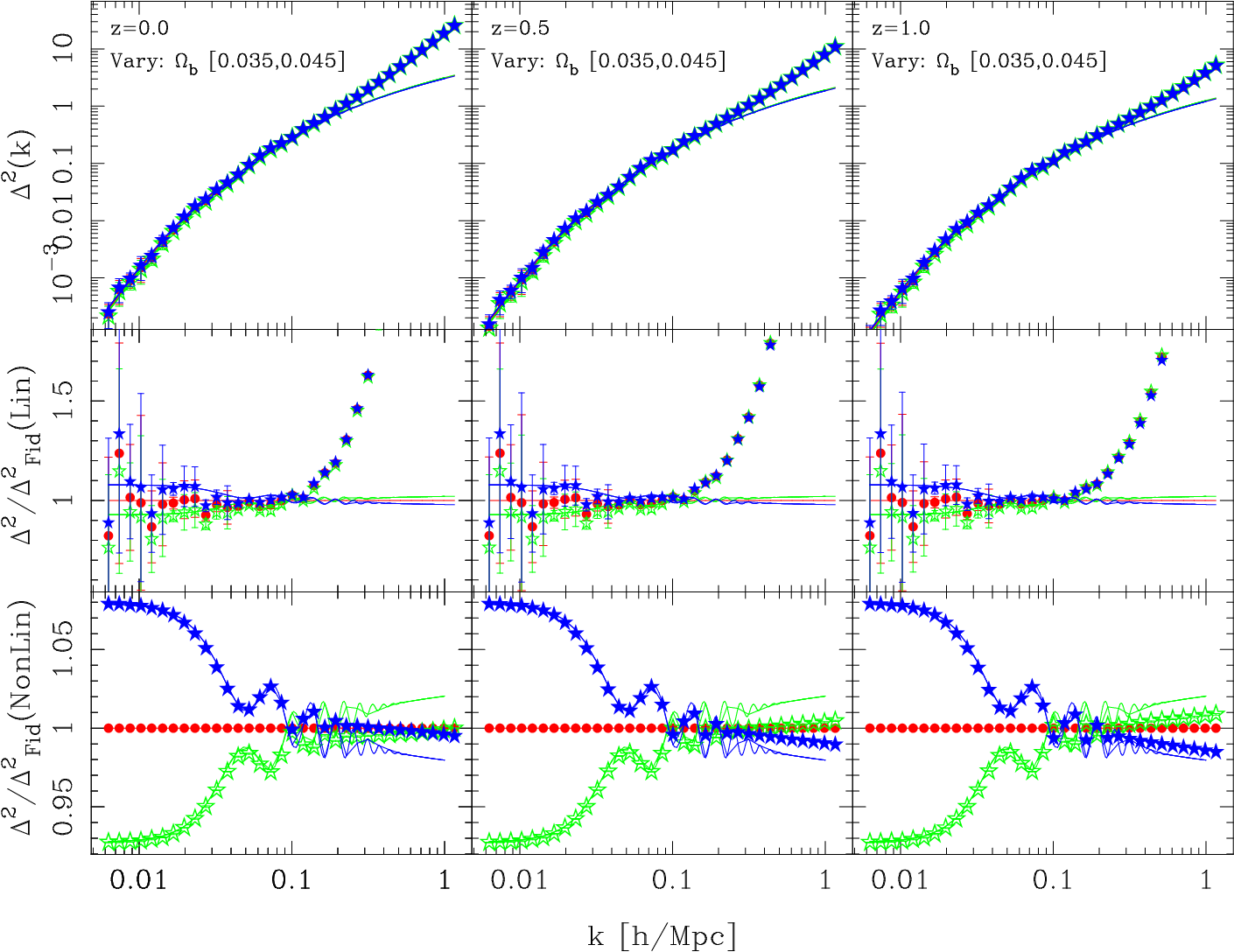}}
\caption{\small{The same as \Fig{fig:PowVar1}, only this time showing
    the dependence of the power spectrum on the baryon density
    parameter $\Omega_b$. The solid blue and open green stars depict
    the results for $\Omega_b=\{0.045,0.035\}$, and the solid red
    points denote the results for the fiducial model
    $\Omega_m=0.04$.\label{fig:PowVar3}}}
\centering{
\includegraphics[width=10cm,height=13cm,angle=-90.0,clip=]{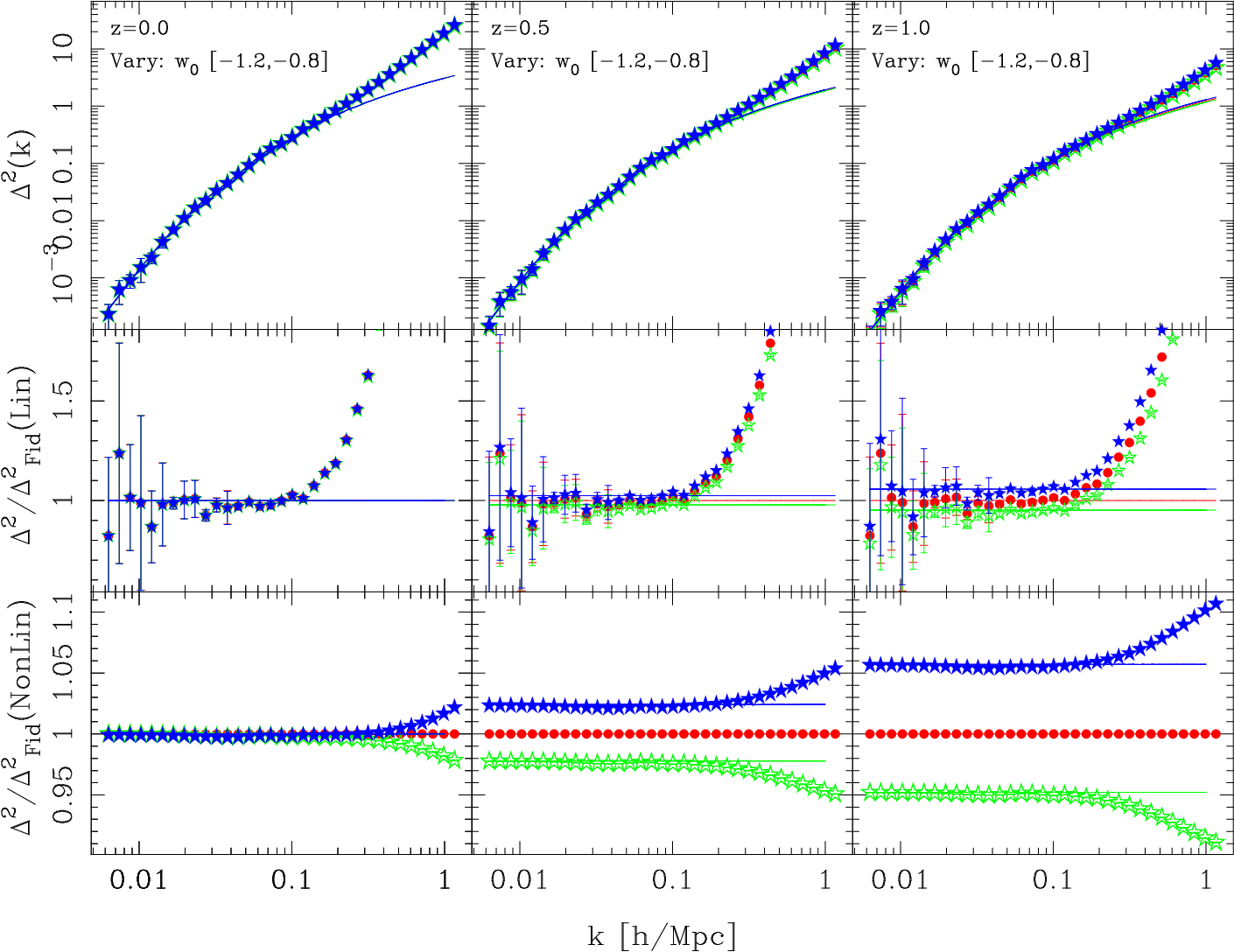}} 
\caption{\small{The same as \Fig{fig:PowVar1}, this time showing the
    dependence of the power spectrum on the dark energy
    equation-of-state parameter $w_0$. The solid blue and open green
    stars depict the results for $w_0=\{-0.8,-1.2\}$, and the solid
    red points denote the results for the fiducial model
    $w_0=-1.0$.\label{fig:PowVar4}}}
\end{figure*} 

\begin{figure*}
\centering{
\includegraphics[width=10cm,height=13cm,angle=-90.0,clip=]{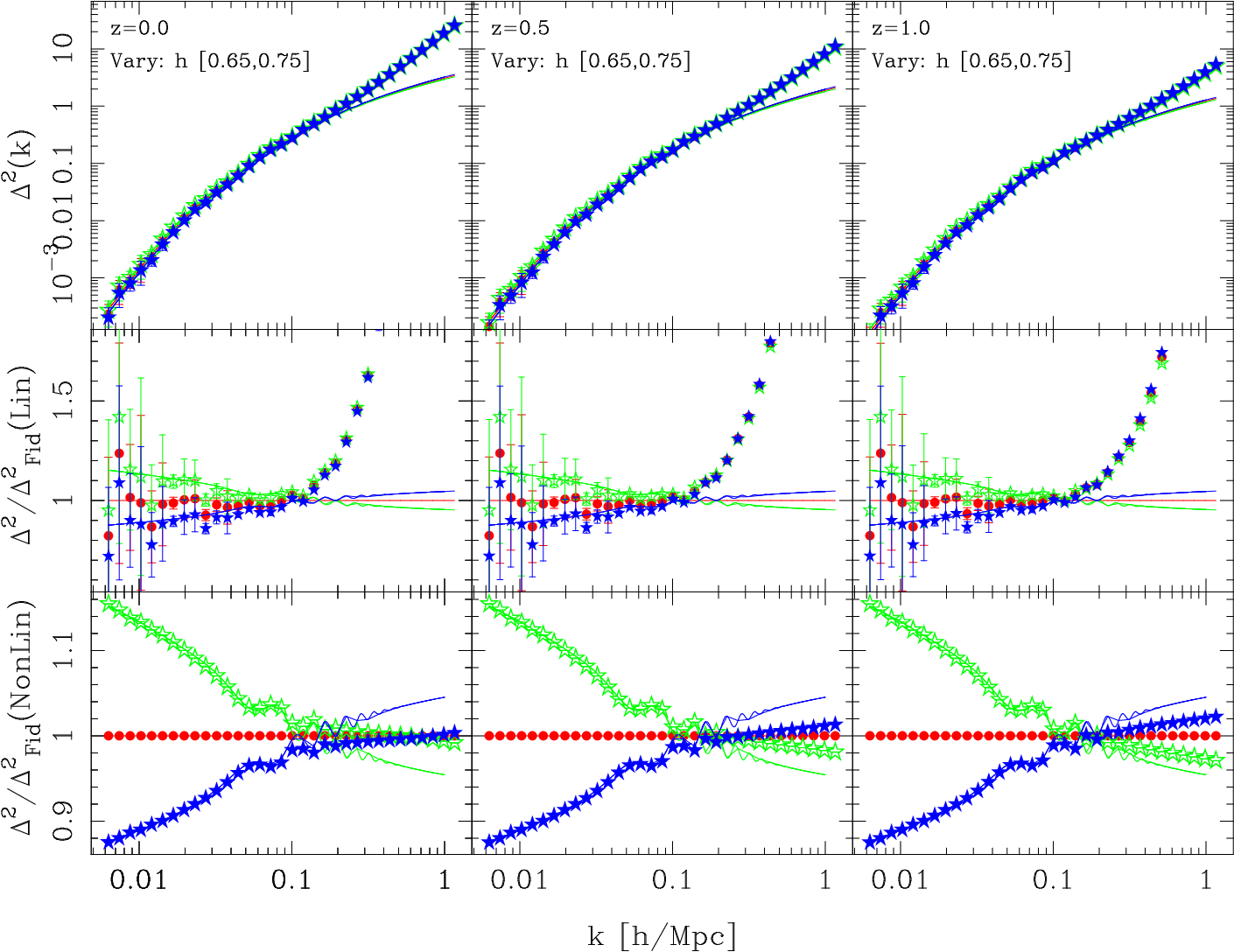}}
\caption{\small{The same as \Fig{fig:PowVar1}, this time showing the
    dependence of the power spectrum on the dimensionless Hubble
    parameter $h$. The solid blue and open green stars depict the
    results for $h=\{0.75,0.65\}$, and the solid red points denote the
    results for the fiducial model $h=0.7$.\label{fig:PowVar5}}}
\vspace{0.5cm}
\centering{
  \includegraphics[width=10cm,height=13cm,angle=-90.0,clip=]{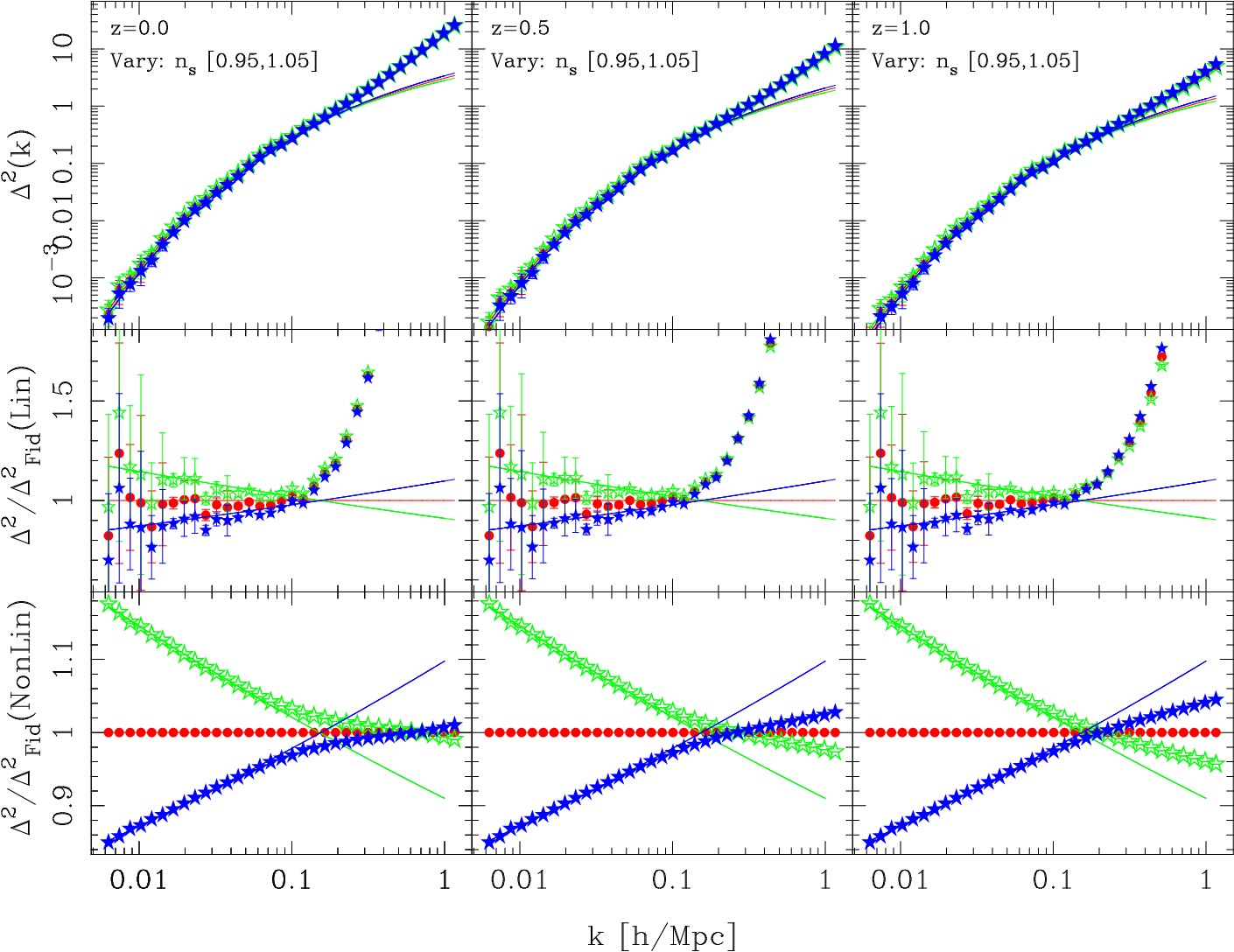}}
\caption{\small{The same as \Fig{fig:PowVar1}, this time showing the
    dependence of the power spectrum on the primordial-power-spectrum
    index $n_s$. The solid blue and open green stars depict the
    results for $n_s=\{1.05,0.95\}$, and the solid red points denote
    the results for the fiducial model $n_s=1.0$.\label{fig:PowVar6}}}
\end{figure*} 

\begin{figure*}
\centering{
\includegraphics[width=10cm,height=13cm,angle=-90.0,clip=]{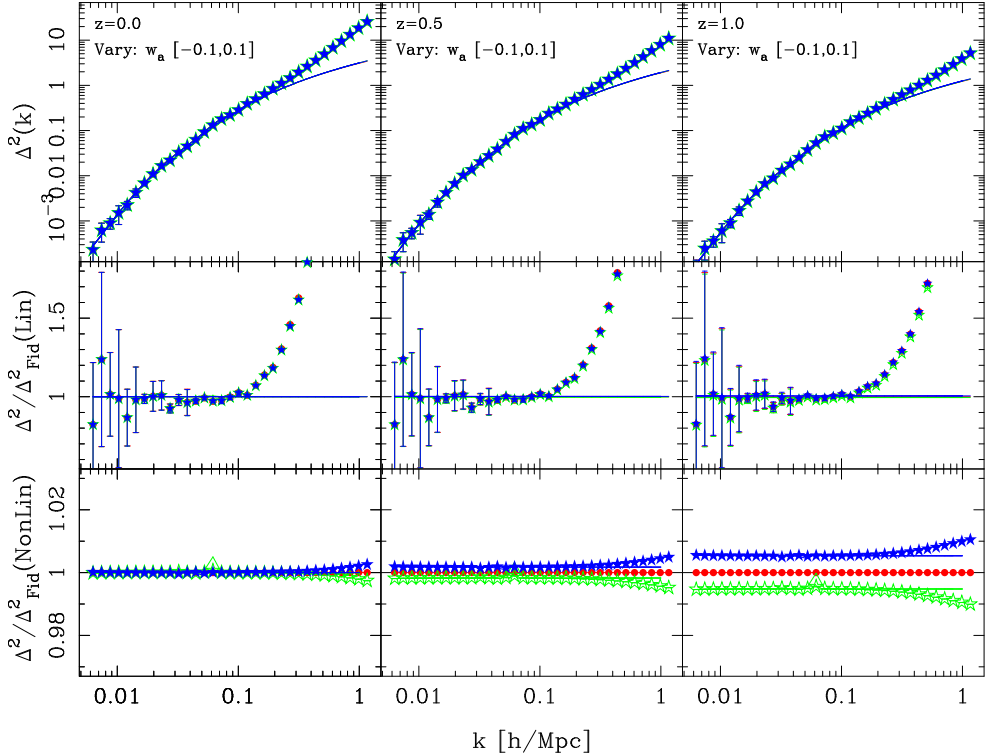}} 
\caption{\small{The same as \Fig{fig:PowVar1}, this time showing the
    dependence of the power spectrum on the dark energy
    equation-of-state parameter $w_a$. The solid blue and open green
    stars depict the results for $w_a=\{0.1,-0.1\}$, and the solid red
    points denote the results for the fiducial model
    $w_a=0$.\label{fig:PowVar7}}}
\end{figure*} 


\section{Analysis II: Variations}\label{sec:variations}


\subsection{Power spectrum dependence on cosmological parameters}

We now turn to the task of exploring the cosmological dependence of
the power spectrum. As mentioned in \S\ref{sec:simulations}, we
consider the variations with respect to 7 cosmological parameters:
\mbox{$\bm\theta=\{\sigma_8,\Omega_m,\Omega_b,w_0,w_a,h,n_s\}$}. For
each cosmological parameter, we freeze all of the other parameters and
simulate two variations, up and down, around the fiducial-model
value. For each such variation we have performed 4 realizations.

Figures~\ref{fig:PowVar1}--\ref{fig:PowVar7} present the
ensemble-averaged variations. For each figure the left, middle and
right panels show the results at redshifts: $z=\{1,0.5,0\}$,
respectively. The top sections show the absolute power; the middle,
the ratio with respect to the linear theory of the fiducial model; and
the bottom the ratio with respect to the measured nonlinear power in
the fiducial model. In all panels the red points denote the fiducial
model, the blue stars denote the upper variation,
i.e. $P(k_i|\bm\phi+\Delta\phi_\mu)$, and the open green stars denote
the lower variation, i.e. $P(k_i|\bm\phi-\Delta\phi_\mu)$. It is worth
noting that when we compute the ratio of the variant power spectra
with the nonlinear fiducial spectrum, we compute this ratio for each
realization, before averaging. This leads to the cancellation of some
of the cosmic variance, and explains why the error bars in the lower
sections of each panel are not visible.

We next turn our attention to the computation of the derivatives of
the power spectra with respect to the cosmological parameters.  In
order to obtain low-noise estimates of these, we take advantage of the
matched initial conditions between the variations and use the
double-sided derivative estimator,
\be
 \widehat{\frac{\partial P(k|\bm\phi)}{\partial \phi_{\alpha}} } = 
 \widehat{P}(k|\bm\phi) \widehat{\frac{\partial \log P(k|\bm\phi)}{\partial\phi_{\alpha}}}\ ,
\ee
where for the first term on the right-hand-side we take all 200 of the
fiducial simulations as described by \Eqn{eq:powest}. For the
logarithmic derivative we use the estimator:
\ba
\widehat{\frac{\partial \log P(k_i|\bm\phi)}{\partial\phi_{\mu}}} & = &  
\frac{1}{N^{\rm var}_{\rm ensemb}}
\sum_{\alpha=1}^{N^{\rm var}_{\rm ensemb}} \nn \\
& & \hspace{-1.5cm} \times 
\left[ \frac{P^{(\alpha)}(k_i|\bm\phi+\Delta\phi_\mu)-P^{(\alpha)}(k_i|\bm\phi-\Delta\phi_\mu)}
{2\Delta\phi_\mu P^{(\alpha)}(k_i|\bm\phi)} \right]
\label{eq:logderiv}\ea
where $N^{\rm var}_{\rm ensemb}=4$. 

\begin{figure*}
\centering{ 
\includegraphics[width=13cm,height=21.5cm,clip=]{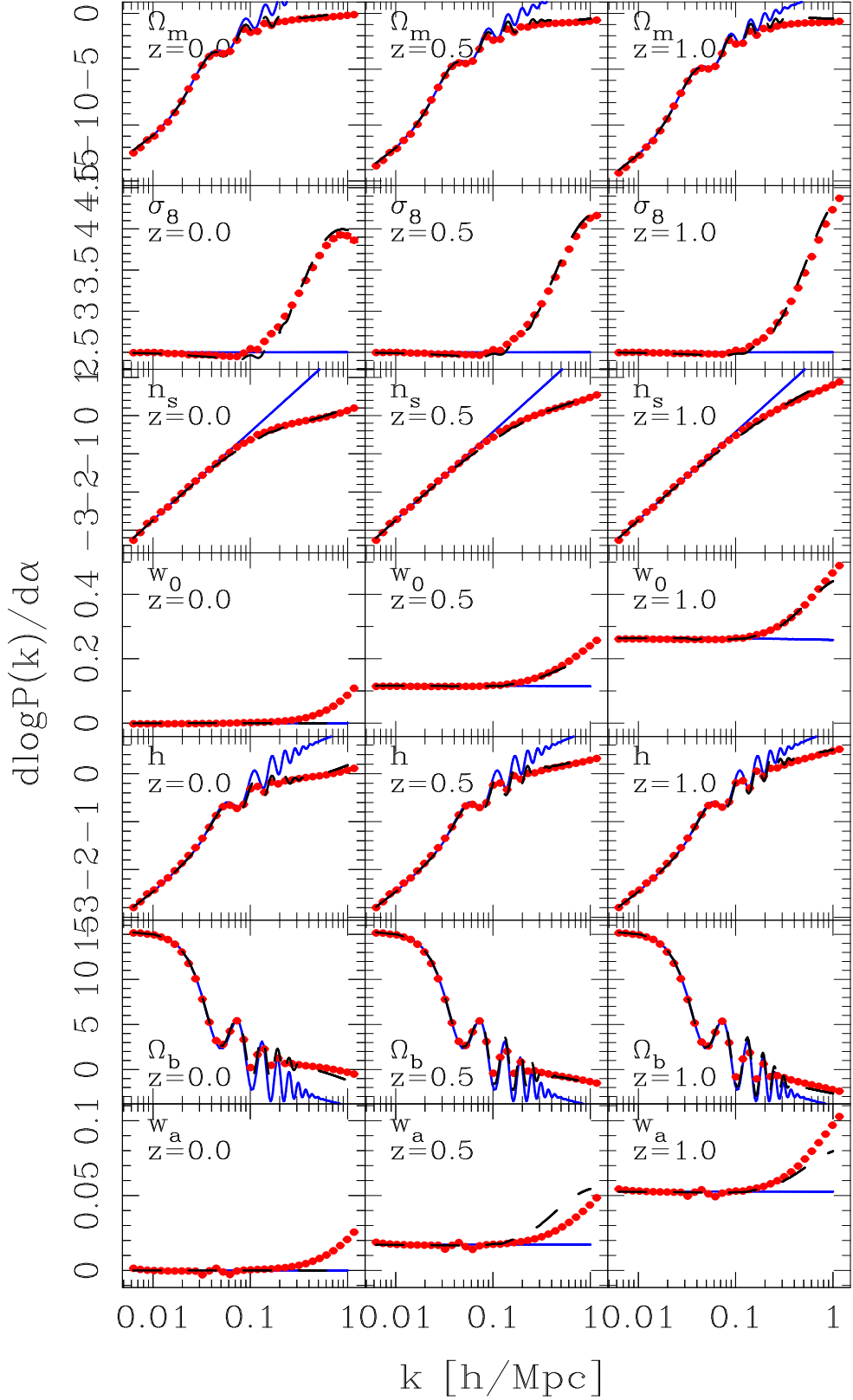}}
\caption{\small{Evolution of the logarithmic derivatives of the power
    spectrum with respect to the cosmological parameters. In all
    panels: solid red points denote the estimates from the $N$-body
    simulations; solid blue and black dashed lines denote the
    predictions from the linear theory and nonlinear {\tt halofit}
    fitting function. From top to bottom the different rows show the
    results for variations in the parameters:
    $\{\Omega_m,\sigma_8,n_s, w_0, h, \Omega_b\}$, respectively. The
    left, central and right columns show the results for epochs,
    $z=\{0, 0.5, 1\}$, respectively.}
\label{fig:FisherDeriv}}
\end{figure*}

Figure~\ref{fig:FisherDeriv} presents the evolution of the logarithmic
derivatives of the power spectrum with respect to the 7 cosmological
parameters that we have considered. The derivatives are computed as
described by \Eqn{eq:logderiv}. In each panel the linear-theory
derivatives are given by the solid blue lines and the black dashed
lines show the predictions for the derivatives using {\tt halofit}
\citep{Smithetal2003}. 

The figure demonstrates that on scales $k<0.1\kMpc$ one may capture
the cosmological parameter dependence of the matter power spectrum
through the variations in the linear power spectrum. However, on
smaller scales on must obtain the derivatives from full nonlinear
modelling. With the exception of $w_0$ and $w_a$ at $z=0$, the
predictions from \mbox{{\tt halofit}} are in reasonably good agreement
with the estimates from the numerical simulations. For $w_0$ and $w_a$
at $z=0$ {\tt halofit} fails to predict the nonlinear
derivatives. This may partially be explained by the fact that we have
normalized the initial power spectra to have the same $\sigma_8$: had
we instead adopted $A_s$, the amplitude of the primordial power
spectrum, as our power spectrum normalization criterion, then we
expect that {\tt halofit} would have made more reasonable predictions
\citep{Jenningsetal2010}.

We also point out that as $k\rightarrow1\kMpc$, the measured
derivatives for $\{\Omega_m,\Omega_b,n_s,h\}$ appear to approach
$\partial \log P/\partial \alpha\rightarrow 0$. This suggests that
there is very little cosmological information to be gained by the
inclusion of measurements on very small scales.  On the other hand,
including the information from small scales can greatly increase the
cosmological information about the parameters $\{\sigma_8,w_0,w_a\}$.


\begin{figure*}
\centering{
\includegraphics[width=12cm,angle=-90.0,clip=]{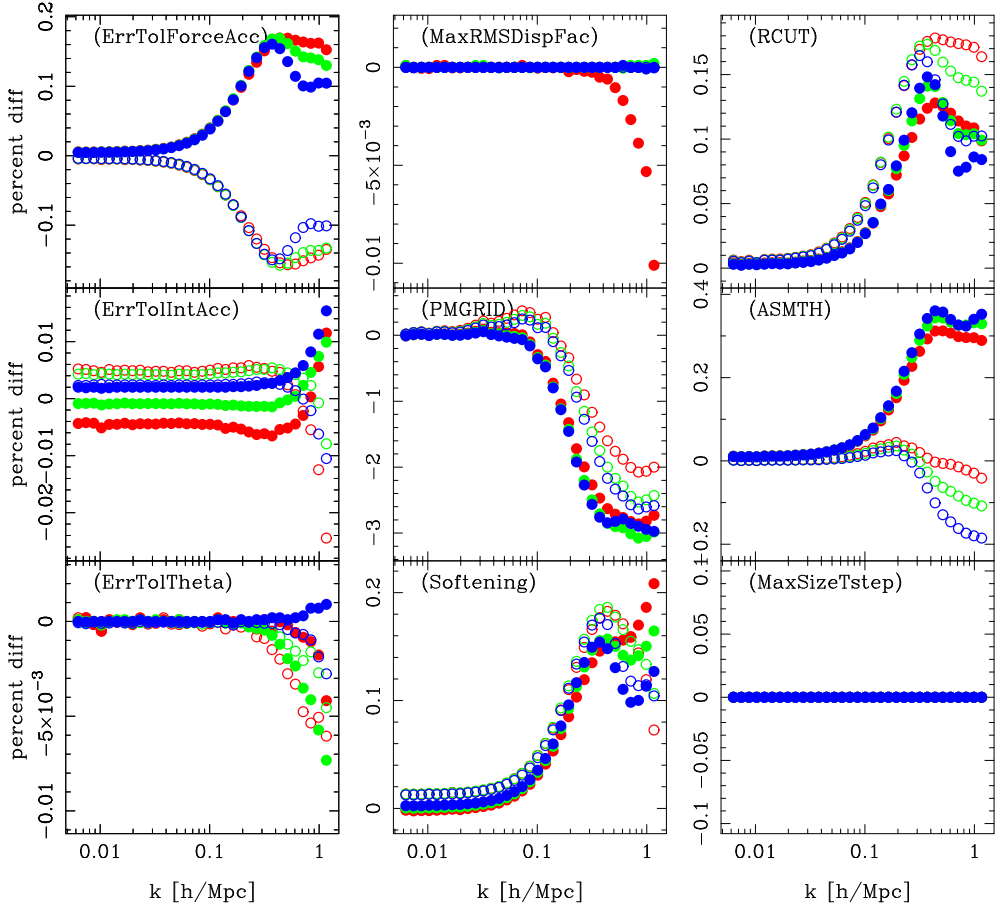}}
\caption{\small{Percentage difference in the simulated power spectra
    as a function of wavenumber. We plot \mbox{$\left[P(k|\psi_{\rm
          fid}\pm\Delta\psi)-P(k|\psi_{\rm fid})\right]/P(k|\psi_{\rm
        fid})$}. Blue, green and red points are for
    $z=\{0,0.5,1\}$, respectively. The open and solid points are
    for the positive and negative changes in the fiducial parameters.}
\label{fig:PowSim}}
\end{figure*}


\subsection{Power spectrum dependence on simulation parameters}

We now explore the dependence of the matter power spectrum on the {\tt
  Gadget-2} simulation parameters. As described in
\S\ref{sec:simulations} we have considered variations in 9 of the
parameters. As for the variations in the cosmological parameters, we
take upper and lower variations of a single simulation parameter and
freeze all others at their fiducial values.

Figure~\ref{fig:PowSim} presents the percentage differences between
the variational and the fiducial models. Each of the nine panels
corresponds to one parameter, with the solid and open points depicting
the upper and lower variations respectively. The blue, green and red
coloured symbols show the results at epochs $z=\{0,0.5,1\}$,
respectively. We find that the most significant source of error is
given by the parameter {\tt PMGRID}, which can yield percent-level
errors on small scales.  We also find that the parameters which
control the interpolation between the Tree- and the PM-force
calculations, {\tt RCUT} and {\tt ASMTH}, can also introduce
significant, but sub-percent errors. Again, these are most important
on small scales. {\tt ErrTolForceAcc} and the {\tt Softening} can also
induce $\sim0.2\%$ errors in the power spectrum. 

It is interesting to note that for the case of the parameters {\tt
  ErrTolForceAcc}, {\tt ErrTolIntAcc}, {\tt ASMTH} the differences
with respect to the fiducial model are almost symmetric for the
positive and negative parameters steps. This suggests that these
parameters are not at their converged values -- it is likely that
decreasing {\tt ErrTolForceAcc} and {\tt ErrTolIntAcc} will always
lead to improved results since they control the accuracy of the
integration. On the other-hand the parameters {\tt RCUT}, {\tt
  PMGRID}, and the {\tt Softening} are not symmetric -- this suggests
that it is not so easy to understand how these parameters affect the
accuracy of the simulations. For the case of {\tt PMGRID} we speculate
that there may be issues associated with beat coupling between the
initial particle lattice -- the memory of which is not lost until late
times -- and the Fourier mesh used to solve the Poisson equation.  For
the fiducial case, since the number of particles and the PMGRID were
identical this effect would be minimised. As one moves to a different
mesh then this effect occurs and induces an error which depends only
on the absolute step size. For the case of the force softening it is
well known that if one uses a softening that is either too large or
too small, then the structure on the smallest scales can be damped.
For the case of too small softening this occurs because hard two-body
encounters can eject particles more easily from potential wells.  In
the case of Fig.~\ref{fig:PowSim} we see that the power increases for
both positive and negative steps. This might be explained by the fact
that if the inner-densities are decreased, then the outer edges of
clusters will have an increased power amount of matter and hence an
increased power spectrum. Indeed our plots do indeed show a turnover
at $k\sim0.5\kMpc$ -- the turn up at higher $k$ may be due to the fact
that the shot noise has not been subtracted. 

\begin{figure*}
\centering{
\includegraphics[width=12cm,angle=-90.0,clip=]{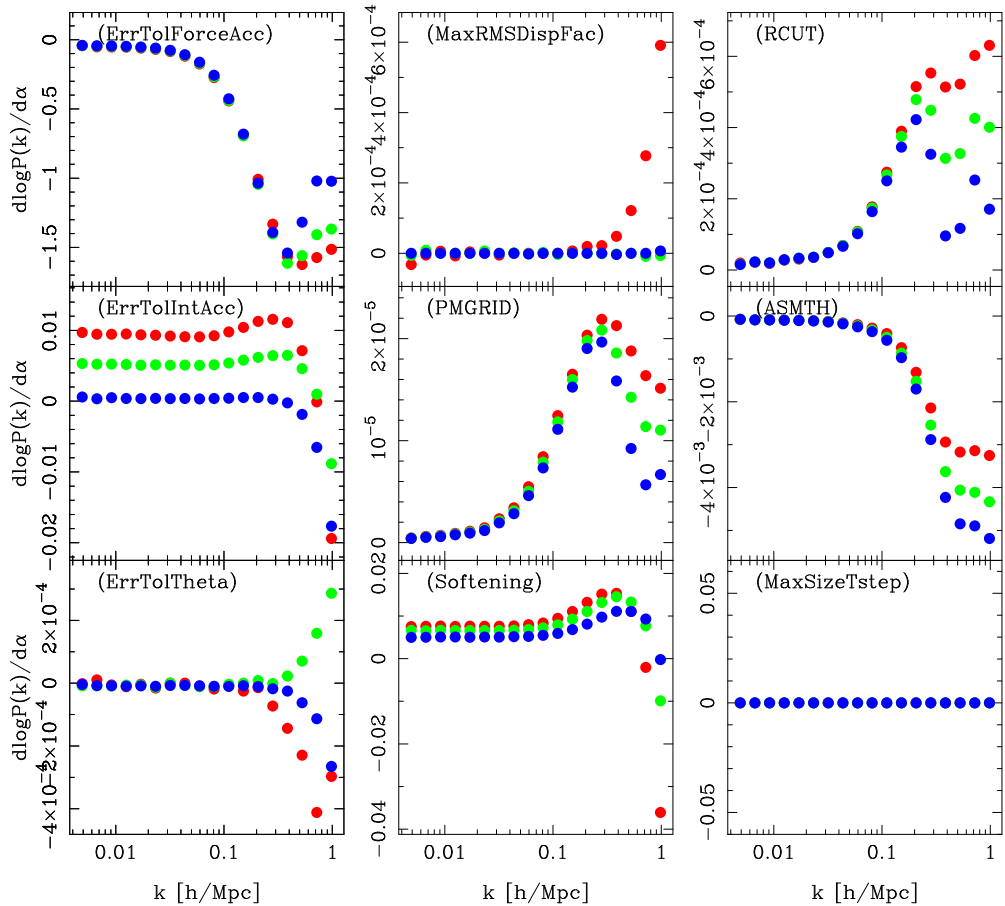}}
\caption{\small{Logarithmic derivatives of the power spectra with
    respect to the simulation parameters as a function of
    wavenumber. Blue, green and red points are for $z=\{0,0.5,1\}$,
    respectively. }
\label{fig:PowSimDiff}}
\end{figure*}

Another important point to note is that the figure also shows that all
of the variations are relatively time-independent. This can be
demonstrated more clearly by considering the logarithmic derivatives.
Figure~\ref{fig:PowSimDiff} presents the logarithmic derivatives of
the matter power spectrum with respect to variations in the {\tt
  Gadget-2} simulation parameters. We estimate the derivatives as
described by \Eqn{eq:logderiv}, except that we only use a single
realization to do this. We now make some important observations:
firstly, on large scales, with the exception of the parameter {\tt
  ErrTolIntAcc}, all of the derivatives are very close to
zero. Moreover, they display a very weak dependence on redshift, which
is a marked difference from the cosmological parameters, which tend to
evolve with both time and scale.  This is an important point,
suggesting that the information coming from the cosmology dependence
of the simulations can be disentangled from that coming from the
simulation parameters.


\section{Fisher Matrix Results}\label{sec:Fisher}

Having obtained estimates for the time evolution of the fiducial model
power spectrum, its covariance matrix and its derivatives with respect
to the cosmological and simulation parameters, we are now in a position
to explore the true cosmological information content of the power
spectrum.


\subsection{Estimator for the Fisher matrix}

We compute the Fisher matrix as described by \Eqn{eq:Fisher}, and
after dropping the first term, the estimator is:
\ba \widehat{\mathcal F}_{\alpha\beta} = \sum_{i,j} 
\frac{ \widehat{\partial \log P_i}}{\partial \phi_\alpha}
 \widehat{P}_i \widehat{C^{-1}}_{ij} \widehat{P}_j
\frac{\widehat{\partial \log P_j}}{\partial \phi_\beta}
\label{eq:Fisher2}\ .
\ea
We shall not use the above equation directly, but an alternate form.
Consider the matrix $\mathbfss{C}$, and let us rewrite it as
\be C_{ij} = \sigma_i \sigma_j r_{ij} \ee
(no summing over repeated indices), where $\sigma^2_i$ is the variance
associated to the $i$th measurement bin. The inverse of $\mathbfss{C}$
can be written as
\be C^{-1}_{ij} = r^{-1}_{ij}/\sigma_i \sigma_j \ .\ee
%
%
Using the above identity allows us to rewrite \Eqn{eq:Fisher2} as:
\ba \widehat{\mathcal F}_{\alpha\beta} = \sum_{i,j} 
\frac{ \widehat{\partial \log y_i}}{\partial \phi_\alpha}
 \widehat{y}_i \widehat{r^{-1}}_{ij} \widehat{y}_j
\frac{\widehat{\partial \log y_j}}{\partial \phi_\beta}
\label{eq:Fisher3} \ ,
\ea
where $\widehat{y}_i\equiv \widehat{y}(k_i|\bm\phi)\equiv
\widehat{P}(k_i|\bm\phi)/\sigma(k_i)$ and
$\sigma(k_i)=\widehat{C}_{ii}^{1/2}$. This latter form is very useful,
since one simply needs to invert the correlation matrix rather than
the covariance matrix. In theory there should be no difference between
the results from these two approaches, however, on a computer there can
be. Whilst the elements of the covariance matrix can differ wildly,
even by orders of magnitude, the elements of the correlation matrix
are constrained to range from $[-1,1]$. Thus we reduce the risk of
inaccurate and potentially unstable inverse estimates due to round-off
errors. This is especially true when large dynamic ranges are
considered and when many matrix elements are employed.


\begin{figure*}
\centering{
\includegraphics[width=7.5cm,angle=-90.0,clip=]{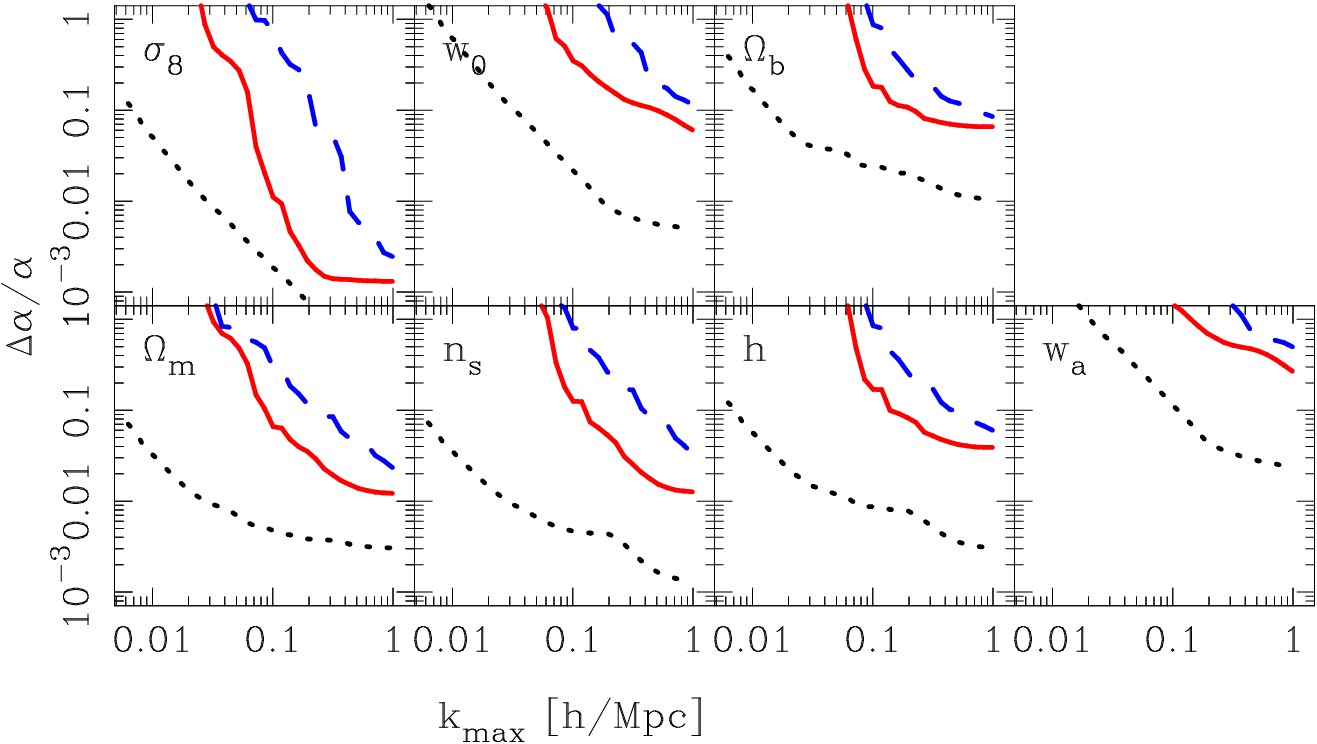}}
\caption{\small{Forecasted fractional 1-$\sigma$ errors on the
    cosmological parameters as a function of the maximum wavenumber
    considered. In all panels, the dashed black lines denote the
    unmarginalized errors; solid red lines denote the errors
    marginalized over all remaining cosmological parameters; the blue
    dashed lines denote the errors after marginalizing over all
    other cosmological parameters and all simulation parameters. The
    top left through to top right panels present the results for
    $\alpha=\{\sigma_8,w_0,\Omega_{\rm b}\}$ and the bottom left to
    bottom right panels present the results for $\alpha=\{\Omega_{\rm
      m},n_s,h,w_a\}$. Note that for $w_a$ we simply plot $\Delta
    w_a$.}\label{fig:1DFisher}} \centering{
  \includegraphics[width=13cm,angle=0.0,clip=]{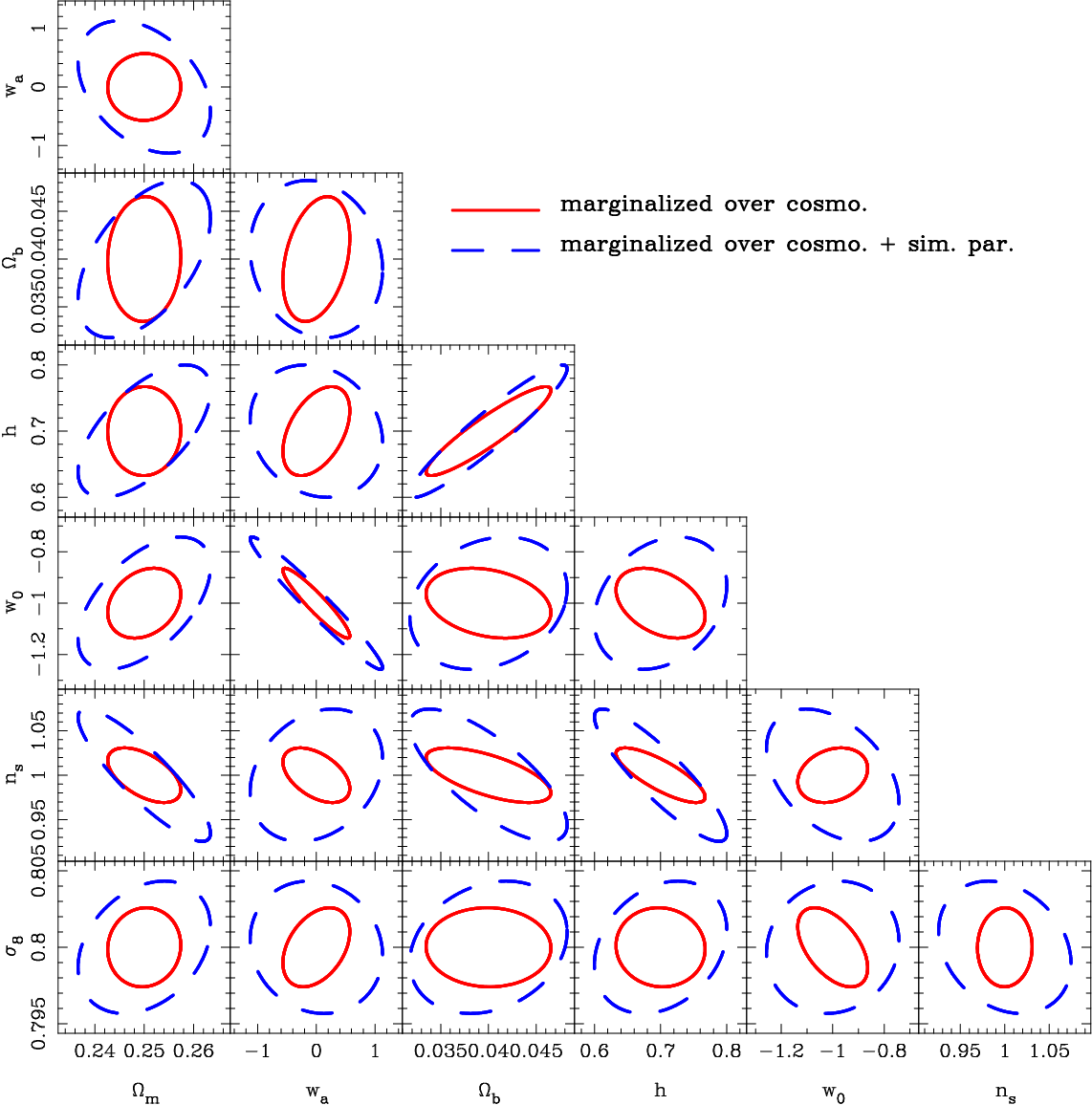}}
\caption{\small{Forecasted likelihood contours for the 15 possible
    parameter combinations. In all of the panels the solid red
    ellipses denote the 1-$\sigma$ likelihood surface one expects from
    our fiducial survey marginalized over all other cosmological
    parameters. The blue dashed ellipses denote the same, but this
    time marginalizing over all other cosmological and simulation
    parameters. }
\label{fig:2DFisher}}
\end{figure*}


\begin{figure*}
\centering{
  \includegraphics[width=7.5cm,angle=-90.0,clip=]{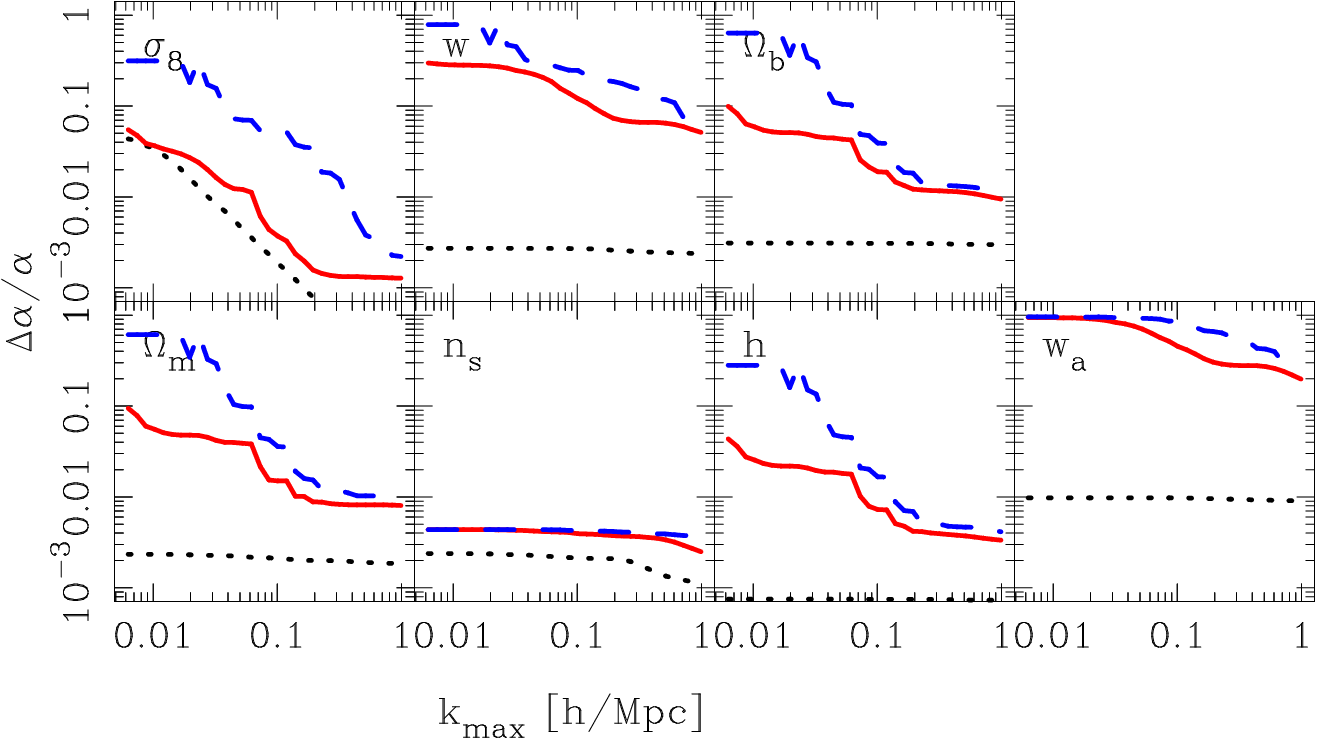}}
\caption{\small{The same as \Fig{fig:1DFisher}, only this time we have
    added the information from a CMB experiment like the Planck
    satellite. Note that we have also employed a strong prior on the
    flatness of the Universe.}
\label{fig:1DFisherPlanck}}
\centering{
\includegraphics[width=13cm,angle=0.0,clip=]{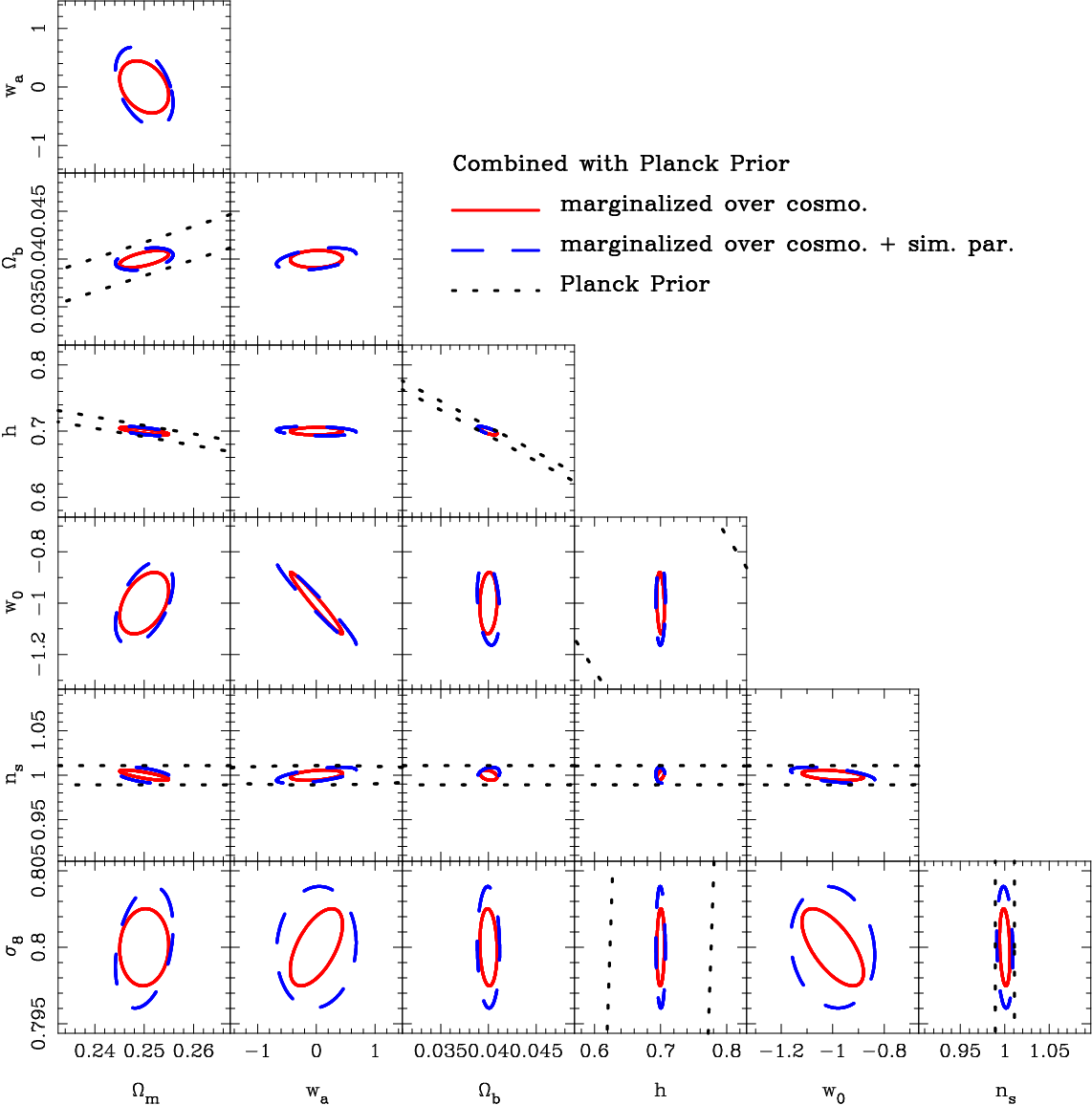}}
\caption{\small{The same as \Fig{fig:2DFisher}, only this time we have
    added the information from a CMB experiment like the Planck
    satellite. Note that we have also employed a strong prior on the
    flatness of the Universe.}
\label{fig:2DFisherPlanck}}
\end{figure*}


\subsection{Information content of the power spectrum}

For our cosmological forecast we adopt a survey consisting of three
independent volumes, each of which has the same volume
$\Vu=3.25\Gpccube$, but mapping the three epochs
$z=\{0,\,0.5,1\}$. Whilst this does not directly match a
particular survey, it covers the scale and evolution that should be
obtainable with BOSS or DES.

Figure~\ref{fig:1DFisher} shows the fractional errors in the matter
power spectrum for the variations in the 7 cosmological parameters, at
epochs $z=\{0,\,0.5,\,1\}$, and as a function of the maximum
wavenumber that enters the calculation, i.e. $k_{\rm max}$.  Note that
since the fiducial value of $w_a=0$, for this case we simply plot
$\Delta w_a$. In the figure, the unmarginalized errors on the
parameters are given by the dotted black lines, i.e.  $\Delta
p_{\alpha} = 1/\sqrt{{\mathcal F}_{\alpha\alpha}}$. This 1-$\sigma$
error is valid only if all the other parameters are known.

If we are required to estimate all parameters from the data then the
best we can ever do is to saturate the MVB, as described in
\S\ref{sec:likelihood}.  If we assume that the simulation parameters
are known, then the errors are given by \Eqn{eq:FishCos}, i.e. $\Delta
p_{\alpha} = \sqrt{\left[{\mathcal F}^{\theta\theta}\right]^{-1}_{\alpha\alpha}}$,
and we obtain the solid red lines in \Fig{fig:1DFisher}. Notice that
the cosmological information is significantly reduced. Once
$k\sim0.4\kMpc$ is reached, adding smaller scales does not reduce the
errors on most of the parameters. This is with the exception of $w_0$,
$w_a$ and $\sigma_8$, for which adding small-scale structures does
help.

On the other hand, if we are to marginalize over the simulation
parameters, owing to the fact that we are ignorant as to the optimal
ones, then the errors are given by \Eqn{eq:FishCosSim}, i.e.  $\Delta
p_{\alpha}=\sqrt{\left[ {\mathcal
      F}^{\phi\phi}\right]^{-1}_{\alpha\alpha}}$. These are
represented in the figure by the blue dashed lines. 

Figure~\ref{fig:2DFisher} shows the 2D likelihood surfaces for
various parameter combinations after marginalizing over all other
parameters. In all of the panels we take $k_{\rm max}=1\kMpc$ and
consider only the 2-$\sigma$ errors, denoted by $\Delta\chi^2=6.17$.
Again the red solid lines present the results for the case where the
simulation parameters are fixed and the blue dashed lines the case
where we marginalize over the simulation parameters. Clearly, there
would be a significant degradation in the constraining power of any
future galaxy clustering survey, should we not be able to identify the
`optimal' simulation parameters. Note also that there appears to be a
strong degeneracy between $\{w_0,w_a\}$ and $\{h,\Omega_b\}$.


\subsection{Combining information from large-scale structures with the CMB}

We now turn to the question of whether adding external data sets may
help alleviate the degradation of the cosmological constraints. Here
we only consider the impact on the errors of adding the information
from a CMB experiment such as Planck \citep{PlanckBlueBook}. Note
that even without Planck data we have already restricted our
exploration of the cosmological parameter space to include only flat
models.  As described in Appendix~\ref{app:planck}, we first compute
the CMB Fisher matrix in a set of parameters that are suitable for the
CMB, and then rotate this matrix to our favoured parameter set for
describing large-scale structure \citep[see also][]{Hilbertetal2012}.
We treat the CMB and Large-scale structure information as independent
and hence the Fisher matrices may be added:
\be 
{\mathcal F}_{\alpha\beta}^{\rm Tot}=
{\mathcal F}^{\rm CMB}_{\alpha\beta}+{\mathcal F}^{\rm LSS}_{\alpha\beta} \ .
\ee

Figure~\ref{fig:1DFisherPlanck} shows again the errors for the 7
cosmological parameters that we have considered. The differences
between the errors obtained from marginalizing over the cosmological
parameters (red lines) and the cosmological-plus-simulation parameters
(blue lines) are significantly reduced.  Thus, inclusion of the CMB
information significantly improves our ability to constrain the
cosmological model.

Figure~\ref{fig:2DFisherPlanck} shows how the 95\%-confidence-level
error ellipses changed when we add the CMB information. We see again
that the constraining power of the combined experiments significantly
improves our ability to constrain cosmology, and also marginalize over
the simulation parameters.


\begin{figure}
\centering{
\includegraphics[width=8cm,angle=0.0,clip=]{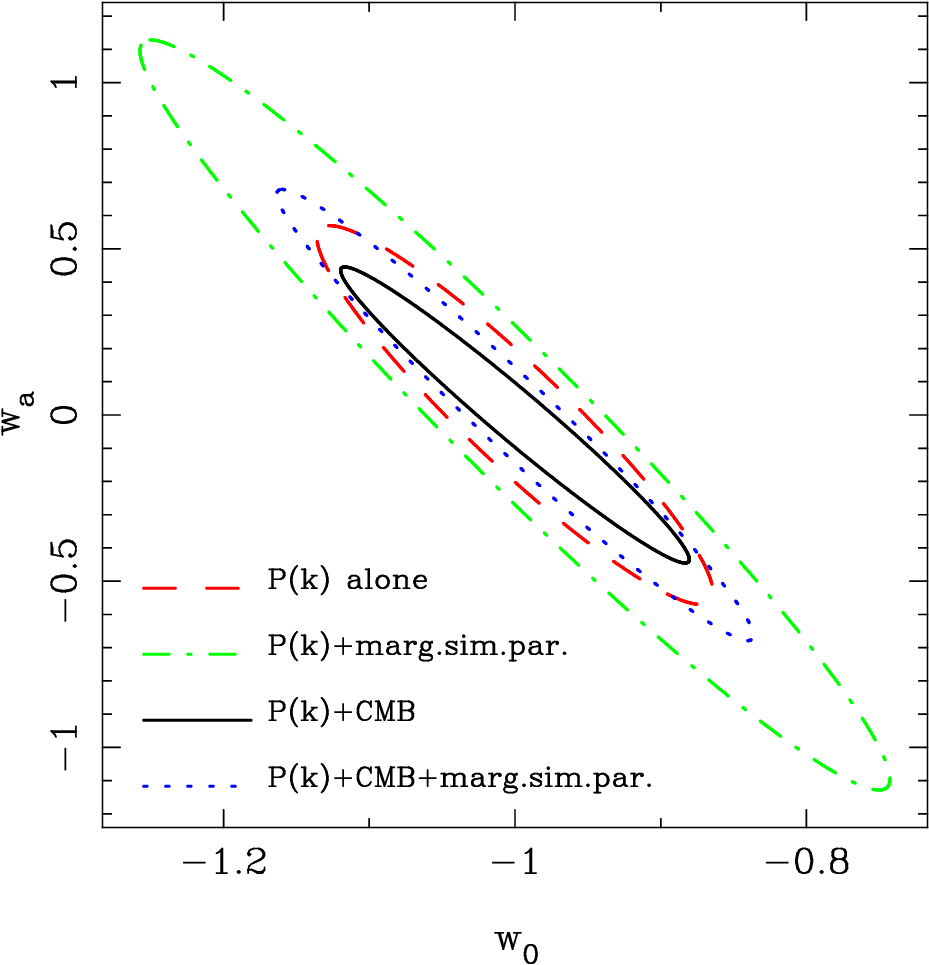}}
\caption{\small{Comparison of the $95\%$-confidence-level contours in
    the likelihood surface of $\{w_0,w_a\}$ after marginalization. The
    red dashed line corresponds to the power spectrum information
    alone, and the green dot-dash line is the same but after
    marginalizing over the simulation parameters. The black solid line
    is the combination of the power spectrum information with a
    Planck-like prior, and the blue dotted line is the same but after
    marginalizing over the simulation parameters.}
\label{fig:2DDarkEnergy}}
\end{figure}


\begin{figure}
\centering{
\includegraphics[width=8cm,angle=0.0,clip=]{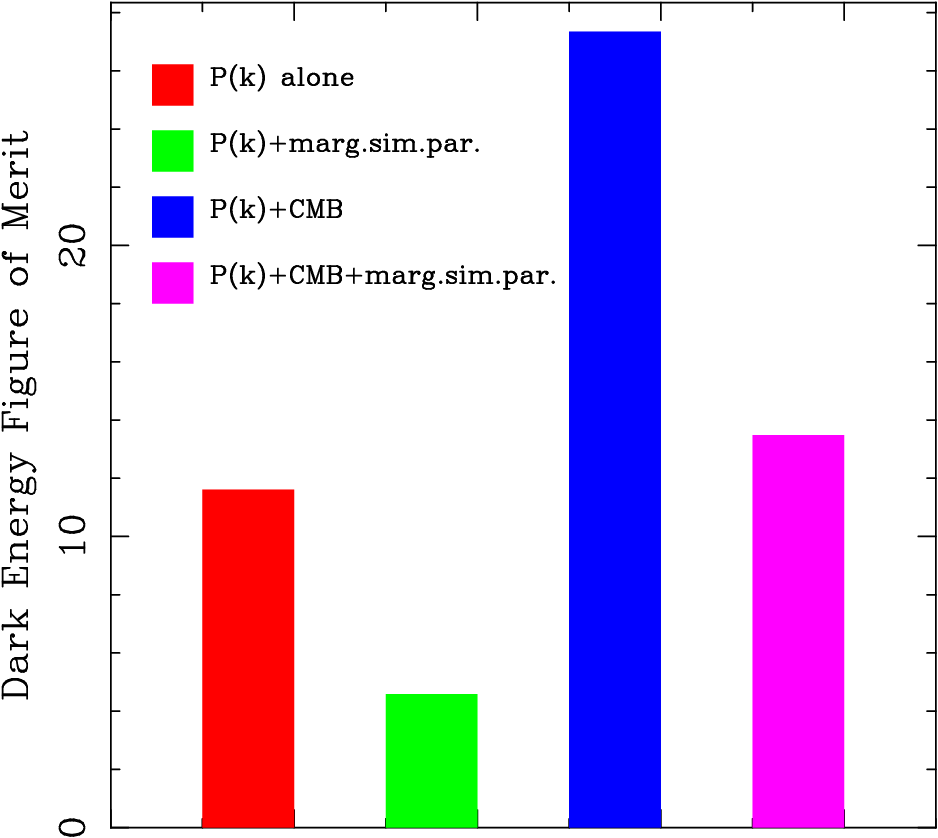}}
\caption{\small{Comparison of the dark energy figures of merit. The
    solid red block represents the case of the power spectrum
    information alone and the green block denotes the same only after
    marginalization over the simulation parameters. The solid blue
    block depicts the case of the power spectrum information combined
    with a Planck-like CMB experiment and the magenta block denotes
    the same, after marginalization over the simulation
    parameters.}
\label{fig:FOM}}
\end{figure}


\section{Constraining Dark Energy}\label{sec:DE}

We now turn to the question of how well we may constrain the time
evolution of the dark energy equation of state, i.e. $\{w_0,w_a\}$.

Figure~\ref{fig:2DDarkEnergy} shows the 95\%-confidence-level
likelihood contours for $\{w_0,w_a\}$, where we use all information
from LSS up to $k=1\kMpc$. The figure reveals that the best
constraints will come from the combination of CMB and LSS
information. However, if we do not understand how to optimize our
$N$-body codes to provide 'optimal' cosmological power spectra, then
marginalizing over the simulation parameters will be costly for Dark
Energy science.

Currently, the standard way to describe the ability of an experiment
to constrain $w(a)$ is through the figure of merit (hereafter FOM)
\citep{DETF2006}.  This has been defined as the inverse area enclosed
by the 2-$\sigma$ error ellipsoid of the parameters $\{w_0,w_a\}$. In
terms of the Fisher matrix, this may be written:
\be {\rm FOM} = \frac{1}{\pi \sqrt{6.17\,{\rm Det[Cov}(w_0,w_a)]}} \ , \ee
where the parameter covariance matrix ${\rm Cov}(w_0,w_a)$ is the
$2\times2$ matrix formed from the submatrix of the $\{w_0,w_a\}$
elements of the inverse of the $7\times7$ Fisher matrix
\citep[e.g. see][]{Wang2008}.

Figure~\ref{fig:FOM} compares the various dark energy figures of
merit. From the figure we clearly see that if one marginalizes over the 
simulation parameters, then there is roughly a factor of 2 penalty in our 
ability constrain the parameters $\{w_0,w_a\}$.


\section{Discussion: Validity of the Fisher matrix approach}\label{sec:discussion}

We now discuss and emphasize some important caveats to our results.

Firstly, let us reexamine the main premise of the paper -- that for a
given simulation code there are parameters that if not optimally
chosen should be marginalized over. One might argue that simulations
do not have `free' parameters, but parameters that simply control the
accuracy of the numerical integration. In principle this is true,
however, for a given $N$-body algorithm it is not clear that a given
code can practically satisfy this ideal statement. Consider for
example the code {\tt Gadget-2}, one might argue that simultaneously
increasing {\tt RCUT} and decreasing {\tt ErrTolTheta} would lead to
increasingly accurate answers -- since in adopting this limit one is
simply going back to computing forces with direct summation -- no tree
-- no PM grid. However, since the tree-force is a monopole expansion
and since the Ewald summation has not been implemented for the
periodic boundary conditions, the code would become less accurate in
the extreme limit of {\tt RCUT$=L$} and {\tt ErrTolTheta=0}. Even if
this was implemented, then solving the forces through direct summation
is not without error, since pair counts for large numbers of particles
will eventually suffer from round-off errors, and {\tt Gadget-2}
stores particle positions as 4-byte floating point numbers. Moreover,
the order in which one takes the force sum -- nearest neighbours first
or distant particles first -- will change the exact value of the
force. We should also add that we want accurate answers from our
numerical code subject to time, memory, cpu and disk usage
constraints. The solution of using direct-particle summation would
obviously fail the time constraint.  Thus the optimization of the code
parameters for {\tt Gadget-2} remains a non-trivial task.

Let us emphasize that we do not expect to have to marginalize over all
simulation parameters when making cosmological inferences with real
survey data -- more simply put, we wish to know what would be the
price one would have to pay if one failed to do the hard work to
establish the `optimal' set of parameters for a given algorithm --
subject to the constraints mentioned above. The Fisher matrix approach
offers a possible route for quickly establishing an answer to this
question. It also enables us to assess at what point one can stop
worrying about systematic errors in the power spectrum due to
uncertainties in certain simulation parameters. For instance if the
degradation in the figure-of-merit comes from a single parameter, then
one can quickly identify that parameter and study its behaviour and so
remove it from the marginalization step. 

Another question mark concerns the use of a Gaussian posterior. It is
clear that the the form of the likelihood function for obtaining a
given set of measurements of the power spectrum is well described by a
Gaussian. If we take the standard assumption of Gaussian initial
conditions, i.e. Fourier modes are Gaussianly distributed, then the
distribution of power in a given mode is exponentially distributed. If
one considers the power spectrum estimator distribution, which
includes the sum of modes in a given $k$-space shell, this is
$\chi^2$-distributed \citep{Takahashietal2009}. In the limit of large
numbers of modes per $k$-space shell, the $\chi^2$-distribution
becomes Gaussian.  Thus it is understood that the likelihood function
should be a multivariate Gaussian. However, where there is room for
debate is in how one makes constraints on the cosmological parameters,
at this point one needs to get an expression for the posterior
probability function. Using Bayes' theorem this is done by multiplying
the likelihood by the parameter priors. Two options are possible:
uninformative flat priors or if one has detailed knowledge of the
system then one can write down informed priors. Since we wanted to be
conservative we adopted uninformative priors. It is here where further
discussion could be had, since one might argue that the parameter
priors are better known. This is probably the case, however, it is
worth being pessimistic at first. The functional form of the
informative priors is not clear. In some cases the form of the priors
is irrelevant, since as we have shown with sufficiently good data sets
one can constrain certain parameters very well and so break
degeneracies, e.g. $n_s$ is very well determined when galaxy
clustering is combined with the CMB data.  On the other-hand the
choice of priors will most likely matter for inferences concerning the
dark energy parameters.


\section{Conclusions}\label{sec:conclusions}

In this paper we have used a large ensemble of $N$-body simulations to
explore the cosmological information content of the matter power
spectrum. We have also explored how the cosmological information is
degraded when we are uncertain as to what the `optimal' $N$-body
simulation parameters are.

In \S\ref{sec:likelihood} we introduced the `Gemeinsam' likelihood
function, which takes into account the dependence of the theoretical
model on the cosmological and $N$-body simulation parameters.  We
reviewed the Fisher matrix formalism for forecasting constraints
obtainable from measurements of the matter power spectrum. The
constraints required us to estimate the fiducial model power spectrum,
its covariance matrix and the first order derivatives of the matter
power spectrum with respect to the cosmological and simulation
parameters. Our fiducial survey consisted of three independent volumes,
each of which had a volume $\Vu=3.25\Gpccube$ but spanning the
redshifts $z=0,\,0.5$ and 1.

In \S\ref{sec:simulations} we described the large ensemble of
simulations that we have performed in order to compute the Fisher
matrix.  We ran 200 simulations to generate the covariance matrix, we
ran 56 simulations to explore the variations of the power spectrum
with respect to the cosmological parameters; and 18 simulations to
explore its dependence on the simulation parameters.

In \S\ref{sec:fiducial} we presented the results for the fiducial
model. We demonstrated that, for $k<0.1\kMpc$ the errors in the power
spectrum were reasonably well described by the Gaussian
prediction. However, on smaller scales the errors were found to be
significantly larger, and were consistent with the presence of a
connected trispectrum that scaled as $T_{ii}\propto P_i^2$. We
explored the off-diagonal covariance of the power spectrum and found
that different band powers were $>50\%$ correlated for $\{k_i,k_j\}\ge
0.1\kMpc$ at $z=0$, and for $\{k_i,k_j\}\ge 0.2\kMpc$ by $z=1$.  We
conclude that non-trivial band power correlations must be accounted
for in the cosmological analysis of large-scale structure data.

In \S\ref{sec:variations} we computed the logarithmic derivatives of
the power spectrum with respect to 7 of the cosmological parameters:
\mbox{$\bm\theta=\{\sigma_8,\Omega_m,\Omega_b,w_0,w_a,h,n_s\}$}. We
found that for $k<0.1\kMpc$ the cosmological dependence could be
reasonably well-captured through the variations in the linear theory
spectra. On smaller scales, the measurements showed strong departures
from the linear predictions. Interestingly, we found that for the
parameters $\{\Omega_m,\Omega_b,n_s,h\}$, and at late times, $\partial
\log P/\partial \alpha\rightarrow 0$, as $k\rightarrow1\kMpc$. This
suggested that there may be very little additional cosmological
information to be gained on these parameters by the inclusion of
measurements on very small scales.  However, we also showed that for
$\{\sigma_8,w_0,w_a\}$ the inclusion of small scales significantly
increases the cosmological information about these parameters.

We then explored the dependence of the matter power spectrum on 9 of
the simulation parameters used for the {\tt Gadget-2} code. We found
that variations in the choice of {\tt PMGRID}, {\tt RCUT}, {\tt
  ASMTH}, {\tt ErrTolForceAcc} and {\tt Softening} could in combination
lead to percent-level variations in the power spectrum on small
scales.
 
In \S\ref{sec:Fisher} we used the simulations to explore the
cosmological information content of the matter power spectrum. We
found that, under the assumptions of flat cosmological models, our
fiducial survey could constrain $\{\sigma_8,\Omega_m,n_s\}$ at the
percent level or better, $\{\Omega_b,h,w_0\}$ at the few-percent
level and $w_a$ at the 20\% level. However, if we fold into our
likelihood analysis uncertainties in the simulation parameters then
all of these constraints are degraded by roughly a factor of 2.  We
then showed that adding external data sets, such as a Planck-like CMB
survey, can help to mitigate the effects of marginalization over the
simulation parameters. In particular, the parameters
$\{n_s,h,\Omega_b,\Omega\}$ are almost unaffected by the
marginalization procedure.

In \S\ref{sec:DE} we focused on the dark energy equation of state
parameters $\{w_0,w_a\}$. We have shown that marginalizing over the
simulation parameters significantly degrades our ability to constrain
the Dark Energy from the power spectrum. Adding the CMB information
does help somewhat. However, we have computed the dark energy figure
of merit and found that there is a factor of $\sim2$ degradation
when the simulation parameters are marginalized over.

In this paper we have worked with the simulation code {\tt Gadget-2}
and a sub-set of parameters that are specific to it. As discussed in
\citet{Reedetal2012}, accurate simulating of cosmic structure
formation involves more than the code parameters. We have neglected to
explore the dependence of the information on the number of simulation
particles, the box-size, the initial start redshift. Thus, taken at
face value this appears to be an optimistic assessment of the problem.

On the other hand, in principle, a number of the issues raised in this
paper might be mitigated by larger simulations: e.g., if one increased
the number of particles $N$ without limit, then the scale at which the
force softening modifies the results could be pushed to higher
wavenumber, since $k_{\rm soft}\equiv 2\pi/l_{\rm soft}\propto
N^{1/3}$. Hence, one could in principle find an $N$ sufficiently large
that $k_{\rm soft}$ will be larger than the targeted wavemodes of the
designed survey. However, finite resources may make this
computationally challenging.

Regarding the generality of our conclusions, one might be concerned
that the point in the simulation parameter space that we adopted as
our fiducial point may bias our results and one might ask: how would
the results change if we adopted another set of fiducial simulation
parameters -- ones closer to the {\em optimal} set? If the likelihood
function does not vary rapidly over the simulation parameter space,
then our estimates for the Hessian and and hence the Fisher matrix
will be robust. This, however, is an important question which will
deserve further attention in the future. We anticipate that answering
it fully will also require one to solve the more subtle problem of
finding the {\em optimal} set of simulation parameters.

This work has also focused on the problem of simulating the dark
matter only power spectrum. Future work will also have to extend this
analysis to include the impact of baryonic physics effects on the
clustering due to: our approximate handling of the evolution of the
coupled baryon-CDM fluid after recombination \citep{SomogyiSmith2010};
uncertainties in the small-scale feedback processes of galaxy
formation \citep{vanDaalenetal2011,Sembolonietal2011}. In addition,
when exploring alternative cosmological models, new uncertainties will
need to be folded into the estimates for example if we also wish to
constrain the dark matter model
\citep{Vieletal2012,Schneideretal2012,Birdetal2012}.


\section*{Acknowledgements}

We would like to thank the anonymous referee for their useful
comments, which helped improve the quality of the draft. We also wish
to thank Pablo Fosalba, Cristiano Porciani, Roman Scoccimarro, Volker
Springel and Simon White for useful discussions. We kindly thank
V.~Springel for making public {\tt Gadget-2}, and R.~Scoccimarro for
making public his {\tt 2LPT} code.  RES acknowledges support from ERC
Advanced grant 246797 `GALFORMOD'. LM was supported by the Deutsche
Forschungsgemeinschaft (DFG) through the grants MA 4967/1-1 and MA
4967/1-2. BM was supported by an SNF grant. MC was supported by the
Ram\`on y Cajal fellowship.


\bibliographystyle{mn2e}
\bibliography{CosmoSimMargConverg.astro-ph.2.bbl}


\appendix

\section{Planck Fisher matrix}
\label{app:planck}

\subsection{Computing the CMB matrix}

In computing the Planck Fisher matrix we follow the methodology
described in \citet{Eisensteinetal1999} and for the specific
implementation we follow \citet{TakadaJain2009}. We thus assume that
the CMB temperature and polarization spectra constrain 9 parameters,
and for our calculations we set their fiducial values to be those from
the recent WMAP7 analysis \citep{KomatsuEtal2011short}. The fiducial
parameters are: dark energy EOS parameters $w_0=-1.0$ and $w_a=0.0$;
the density parameter for dark energy $\Omega_\mathrm{DE}=0.7274$; the
CDM and baryon density parameters scaled by the square of the
dimensionless Hubble parameter $w_\mathrm{c}=\Omega_\mathrm{c}
h^2=0.1125$ and $w_\mathrm{b}=\Omega_\mathrm{b}h^2=0.0226$
($h=H_0/[100 \mathrm{kms^{-1}Mpc^{-1}}]$); and the primordial spectral
index of scalar perturbations $n_\mathrm{s}=0.963$; the primordial
amplitude of scalar perturbations $A_\mathrm{s}=2.173\times10^{-9}$;
the running of the spectral index $\alpha=0.0$; and the optical depth
to the last scattering surface $\tau=0.087$. Hence we may write our
vector of parameters:
\begin{equation}
{\bf p}^T  =  
{(w_0,w_a,\Omega_\mathrm{DE},w_\mathrm{c},w_\mathrm{b},\tau,n_\mathrm{s},A_\mathrm{s},\alpha)}^T.
\end{equation}

The CMB Fisher matrix can be written as \citep{Eisensteinetal1999}:
\def\Fisher{{\mathcal F}}
\begin{equation} \Fisher_{p_\alpha p_\beta}=
\sum_\ell \sum_{X,Y} 
\frac{\partial C_{\ell,X}}{\partial p_\alpha}
\mathrm{Cov}^{-1}\left[C_{\ell,X},C_{\ell,Y}\right]
\frac{\partial C_{\ell,Y}}{\partial p_\beta}\ ,
\end{equation}
where $\{X,Y\}\in \{\mathrm{TT},\,\mathrm{EE},\,\mathrm{TE}\}$, where
$C_{\ell,\rm TT}$ is the temperature power spectrum, $C_{\ell,\rm EE}$
is the E-mode polarization power spectrum, $C_{\ell,\rm TE}$ is the
temperature-E-mode polarization cross-power spectrum. We have been
conservative and assumed that there will be no significant information
from the $C_{\ell,\rm BB}$, the B-mode polarization power spectrum. We
compute all CMB spectra using {\tt CAMB} \citep{Lewisetal1999} and use
the additional module for time evolving dark energy models
\citep{HuSawicki2007}. We include information from all multipoles in
the range ($2< \ell <1500$).

The covariance matrices for these observables are:
\def\cov{{\rm Cov}}
\newcommand{\TT}{{\mathrm{TT}}}
\newcommand{\EE}{{\mathrm{EE}}}
\newcommand{\TE}{{\mathrm{TE}}}
\newcommand{\BB}{{\mathrm{BB}}}
\begin{align} 
\cov\left[C_{\ell,\TT},C_{\ell,\TT}\right] & = \frac{1}{f_\mathrm{sky}}
\frac{2}{2\ell+1}\left[C_{\ell,\TT}+N_{\ell,\TT}\right]^2 \ ;\\ 
\cov\left[C_{\ell,\TT},C_{\ell,\EE}\right] & = \frac{1}{f_\mathrm{sky}}
\frac{2}{2\ell+1}  C_{\ell,\TE}^2 \ ;                \\
\cov\left[C_{\ell,\TT},C_{\ell,\TE}\right] & = \frac{1}{f_\mathrm{sky}}
\frac{2}{2\ell+1} C_{\ell,\TE}\left[C_{\ell,\TT}+N_{\ell,\TT}\right]\ ;\\
\cov\left[C_{\ell,\EE},C_{\ell,\EE}\right] & = \frac{1}{f_\mathrm{sky}}
\frac{2}{2\ell+1} \left[C_{\ell,\EE}+N_{\ell,\EE}\right]^2\ ;\\
\cov\left[C_{\ell,\EE},C_{\ell,\TE}\right] & = \frac{1}{f_\mathrm{sky}}
\frac{2}{2\ell+1}C_{\ell,\TE}\left[C_{\ell,\EE}+N_{\ell,\EE}\right] \ ;\\
\cov\left[C_{\ell,\TE},C_{\ell,\TE}\right] & = \frac{1}{f_\mathrm{sky}}
\frac{1}{2\ell+1} \left[C_{\ell,\TE}^2+\right.\nn \\
 &\quad\left. (C_{\ell,\EE}+N_{\ell,\TT})(C_{\ell,\EE}+N_{\ell,\EE})\right]\ .
\end{align}
In the above $f_{\rm sky}$ is the fraction of sky that is surveyed and
usable for science, and we take $f_{\rm sky}=0.8$.  The terms
$N_{\ell,\TT}$ and $N_{\ell,\EE}$ denote the beam-noise in the
temperature and polarization detectors, respectively. These can be
expressed as:
\begin{align} 
N_{\ell,\TT} & = \left[w_{\TT}W^2_\mathrm{Beam}(\ell)\right]^{-1}\\
N_{\ell,\EE} & =  \left[w_{\EE}W^2_\mathrm{Beam}(\ell)\right]^{-1} \ ,
\end{align}
where
$w_{\TT}=\left[\Delta_\mathrm{T}\theta_\mathrm{Beam}\right]^{-1}$ and
$w_{\EE}=\left[\Delta_\mathrm{E}\theta_\mathrm{Beam}\right]^{-1}$. The
beam window function has the form:
\begin{equation}
W^2_\mathrm{Beam}(\ell)=\exp\left[-\ell(\ell+1)\sigma^2_\mathrm{Beam}\right]; \quad
\sigma_\mathrm{Beam}\equiv \frac{\theta_\mathrm{Beam}}{\sqrt{8\log 2}}\ .
\end{equation}
For the Planck experiment we assume that we have a single frequency
band for science (143 GHz), and for this channel the following
parameters apply \citep{PlanckBlueBook}: angular resolution of the
beam $\theta_\mathrm{Beam}=7.1'$ [FWHM]; the beam intensity is
\mbox{$\Delta_\mathrm{T}=2.2\, (T_\mathrm{CMB}/1\mathrm{K}) \, [\mu\rm
    K]$}, \mbox{$\Delta_\mathrm{E}=4.2\, (T_\mathrm{CMB}/1\mathrm{K})
  \, [\mu\rm K]$}. We take the temperature of the CMB to be $T=2.726
\mathrm{K}$.

\subsection{Transforming from CMB to large-scale structure parameters}

In the formation of the large-scale structure we have considered how
the matter power spectrum depends on the 7 cosmological parameters:
\mbox{$\bm\theta=\{\sigma_8,\Omega_m,\Omega_b,w_0,w_a,h,n_s\}$}
parameters. Let us rewrite our original 9-D CMB parameter set in terms
of a new 9-D large-scale structure parameter set. Let us therefore
consider the transformation:
\ba
{\bf p} & = & {(w_0,w_a,\Omega_\mathrm{DE},w_\mathrm{c},w_\mathrm{b},\tau,n_\mathrm{s},A_\mathrm{s},\alpha)}^T \\
{\bf q} & = & {(w_0,w_a,\Omega_\mathrm{m},h,\Omega_b,\tau,n_\mathrm{s},\sigma_8,\alpha)}^T .
\ea
Five of the parameters are unchanged from the original set.  The
remaining four are related to the original parameters in the following
way:
\ba
\Omega_\mathrm{m} & = & 1-\Omega_\mathrm{DE} \ ;\label{eq:par1}\\
\Omega_b & = & \frac{w_\mathrm{b}}{w_\mathrm{b}+w_\mathrm{c}}(1-\Omega_\mathrm{DE})\ ;\label{eq:par2}\\
h & = & \sqrt{\frac{w_\mathrm{b}+w_\mathrm{c}}{1-\Omega_\mathrm{DE}}}\ ;\label{eq:par3} \\
\sigma_8 & = & \left[\int \frac{\dk}{(2\pi)^3}P(k|{\bf p})
  W(kR)^2\right]^{1/2} , \label{eq:par4}
\ea
where $P(k|{\bf p})$ is the matter power spectrum, which depends on
parameters $p_{\alpha}$, and where the real space, spherical top-hat
filter function has the form $W_k(y)=3[\sin y-y \cos y]/y^3$, with
$y\equiv kR$ and $R=8\Mpc$.

It can be shown that a Fisher matrix in one set of suitable variables
may be represented in another basis space through the transformation:
\be {\mathcal F'}_{\gamma\delta}({\bf q})=\sum_{\gamma,\alpha}\Lambda_{\alpha\gamma}{\mathcal
  F}_{\alpha\beta}({\bf p}) \Lambda_{\beta\delta}\ee
where 
$\Lambda_{\alpha\gamma}\equiv \partial p_{\alpha}/\partial q_{\gamma}$
is the matrix formed from the partial derivatives of the old
parameters with respect to the new ones.  From
Eqns~(\ref{eq:par1})--(\ref{eq:par4}) we have $q_a=G_a({\bf p})$,
however in order to perform the partial derivatives we actually
require the inverse of these relations, i.e.  $p_a=G_a^{-1}({\bf
  q})$. In some cases these inverse relations may easily be
determined, e.g. \Eqn{eq:par1}. However, in other cases no analytic
inverse exists, e.g. \Eqn{eq:par4}.  A simple way around this problem
is through recalling the following:
\begin{equation} 
\sum_{\alpha}\frac{\partial p_{\mu}}{\partial q_{\alpha}}\frac{\partial q_{\alpha}}{\partial p_{\nu}}
= \sum_{\alpha}\Lambda_{\mu\alpha}\Lambda^{-1}_{\alpha\nu}=\delta^{K}_{\mu\nu}
\end{equation}
Hence, if we first compute ${\partial q_{\alpha}}/{\partial
  p_{\nu}}\equiv\Lambda^{-1}_{\alpha,\nu}$, then we may obtain
$\Lambda$, from the fact that: $\Lambda \rightarrow
[\Lambda^{-1}]^{-1}$.

Let us therefore form the matrix $\Lambda^{-1}$. For the case of those
parameters that are unchanged
$\Lambda^{-1}_{\alpha\beta}=\delta^{K}_{\alpha\beta}$. However, for
the remaining ones, we have:
\begin{align}
\frac{\partial \Omega_\mathrm{m}}{\partial \Omega_\mathrm{DE}} & =  -1 \ ; \\
\frac{\partial \Omega_b}{\partial \Omega_{\rm DE}} & =  \frac{-w_\mathrm{b}}{w_\mathrm{c}+w_\mathrm{b}} \ ; \\ 
\frac{\partial \Omega_b}{\partial w_\mathrm{c}} & =  
\frac{-w_\mathrm{b}}{[w_\mathrm{c}+w_\mathrm{b}]^2}(1-\Omega_{\rm DE}) \ ; \\ 
\frac{\partial \Omega_b}{\partial w_\mathrm{b}} & =  
\frac{w_\mathrm{c}}{[w_\mathrm{c}+w_\mathrm{b}]^2}(1-\Omega_{\rm DE}) \ ; \\ 
\frac{\partial h}{\partial \Omega_\mathrm{DE}} & = \frac{1}{2}\sqrt{\frac{w_\mathrm{c}+w_\mathrm{b}}{(1-\Omega_\mathrm{DE})^3}} \ ; \\ 
\frac{\partial h}{\partial w_\mathrm{c}} & =  \frac{1}{2}\left[(1-\Omega_\mathrm{DE})(w_\mathrm{b}+w+c)\right]^{-1/2} \ ; \\
\frac{\partial \sigma_8}{\partial p_{\alpha}} & = 
\frac{1}{2\sigma_8} \frac{\partial \sigma_8^2}{\partial p_{\alpha}} = 
\frac{1}{2\sigma_8} \int \frac{\dk}{(2\pi)^3} \frac{\partial P(k|{\bf p})}{\partial p_{\alpha}} W(kR)^2\ .
\end{align}
Note that in order to compute the derivatives ${\partial
  \sigma_8}/{\partial p_{\alpha}}$ we use the package CAMB. Besides
the generation of various CMB power spectra, this package can output
the present day linear theory matter power spectra $P(k|{\bf p})$.
The derivatives are then determined numerically using the standard
estimator for two sided derivatives. The numerical inverse of the
matrix $\Lambda^{-1}$ can easily be computed using the SVD
algorithm \citep{PressEtal1992}.

\end{document}

%% file: defs.tex
\newcommand{\be}{\begin{equation}}
\newcommand{\ee}{\end{equation}}
\newcommand{\ba}{\begin{eqnarray}}
\newcommand{\ea}{\end{eqnarray}}

\newcommand\Eqn[1]     {Eq.\,(\ref{#1})}

\newcommand\Fig[1]     {Fig.\,{\ref{#1}}}

\newcommand\nn         {\nonumber}

\def\CC{{\rm \bf C}}

\def\pp1{{\prime}}
\def\pp2{{\prime\prime}}

\def\2D{{\rm 2D}}

\def\bx{{\bf x}}

\def\bk{{\bf k}}

\def\1Loop{{\rm 1Loop}}

\def\rhob{\bar{\rho}}

\def\kpc{\, h^{-1}{\rm kpc}}
\def\Mpc{\, h^{-1}{\rm Mpc}}
\def\Mpccube{\, h^{-3} \, {\rm Mpc}^3}
\def\Gpc{\, h^{-1}{\rm Gpc}}
\def\Gpccube{\, h^{-3} \, {\rm Gpc}^3}
\def\kMpc{\, h \, {\rm Mpc}^{-1}}

\def\dx{{\rm d}^3{\!\bf x}}
\def\dk{{\rm d}^3{\bf k}}

\def\nbar{\bar{n}}

\def\fun#1#2{\lower3.6pt\vbox{\baselineskip0pt\lineskip.9pt
        \ialign{$\mathsurround=0pt#1\hfill##\hfil$\crcr#2\crcr\sim\crcr}}}

